\documentclass[11pt,draft]{llncs}
\usepackage{latexsym,amsfonts,amsmath,amssymb}
\advance\hoffset by -1.6cm
\advance\voffset by -0.5cm
\renewcommand{\leq}{\leqslant}
\renewcommand{\geq}{\geqslant}
\newcommand{\CC}{\mathsf{Tree}}
\renewcommand{\tilde}{\widetilde}
\renewcommand{\hat}{\widehat}
\newcommand{\Mor}{\mathrm{Mor}}
\newcommand{\Id}{\mathrm{Id}}
\newcommand{\cami}{\!\rightsquigarrow\!}

\begin{document}

\title{An Algebraic View of the Relation between Largest Common
Subtrees and Smallest Common Supertrees}

\author{Francesc Rossell\'o\inst{1}, Gabriel Valiente\inst{2}}

\institute{Department of Mathematics and Computer Science, Research
Institute of Health Science (IUNICS), University of the Balearic
Islands, E-07122 Palma de Mallorca, \texttt{cesc.rossello@uib.es}\and 
Algorithms, Bioinformatics, Complexity and Formal
Methods Research Group, Department of Software, Technical University of  
Catalonia,
E-08034 Barcelona, \texttt{valiente@lsi.upc.edu}}

\maketitle

\begin{abstract}
The relationship between two important problems in tree pattern
matching, the largest common subtree and the smallest common supertree
problems, is established by means of simple constructions, which allow
one to obtain a largest common subtree of two trees from a smallest
common supertree of them, and vice versa.  These constructions are the
same for isomorphic, homeomorphic, topological, and minor embeddings,
they take only time linear in the size of the trees, and they turn out  
to
have a clear algebraic meaning.
\end{abstract}

% \begin{keyword}
% 
% Tree pattern matching; subtree isomorphism; subtree homeomorphism,
% topological embedding; minor containment; largest common subtree;
% smallest common supertree.
% 
% \end{keyword}

\section{Introduction}

Subtree isomorphism and the related largest common subtree and
smallest common supertree problems have practical applications in
combinatorial pattern
matching~\cite{jiang.ea:jda:2004,pinter.ea:2004,valiente:cpm:jda},
pattern
recognition~\cite{conte.ea:2004,fernandez.valiente:2001,torsello.ea:2005},
computational molecular
biology~\cite{aoki.ea:2003,pinter.ea:2005,zhang:2005}, chemical
structure search \cite{artymiuk.ea:2005,barnard:1993,gillet.ea:2003},
and other areas of engineering and life sciences.  In these areas,
they are some of the most widely used techniques for comparing
tree-structured data.

\emph{Largest common subtree} is the problem of finding a largest tree
that can be embedded in two given trees, while \emph{smallest common
supertree} is the dual problem of finding a smallest tree into which
two given trees can be embedded.  A tree $S$ can be embedded in
another tree $T$ when there exists an injective mapping $f$ from the
nodes of $S$ to the nodes of $T$ that transforms arcs into paths in
some specific way.  The type of embedding depends on the properties of
the mapping $f$.  In this paper we consider the following four types
of tree embeddings, defined by suitable extra conditions on $f$:
\begin{description}
\item [Isomorphic embedding:] if there is an arc from $a$ to
$b$ in $S$, then there is an arc from $f(a)$ to
$f(b)$ in $T$.

\item [Homeomorphic embedding:] if there is an arc from $a$ to
$b$ in $S$, then there is a path from $f(a)$ to
$f(b)$ in $T$ with all intermediate nodes of total degree 2 and no
intermediate node belonging to the image of $f$.

\item [Topological embedding:] if there is an arc from $a$ to
$b$ in $S$, then there is a path from $f(a)$ to
$f(b)$ in $T$ with no intermediate node belonging to the image of $f$;
and if there are arcs from $a$ to two distinct nodes $b$ and $c$ in
$S$, then the paths from $f(a)$ to $f(b)$ and to $f(c)$ in $T$ have no
common node other than $f(a)$.

\item [Minor embedding:] if there is an arc from $a$ to
$b$ in $S$, then there is a path from $f(a)$ to $f(b)$ in
$T$ with no intermediate node belonging to the image of $f$.
\end{description}

The different \emph{subtree embedding problems} of deciding whether a
given tree can be embedded into another given tree, for the different
types of embedding defined above, have been thoroughly studied in the
literature.  Their complexity is already settled: they are
polynomial-time solvable for isomorphic, homeomorphic, and topological
embeddings, and NP-complete for minor  
embeddings~\cite{dessmark.ea:2000,%
kilpeleinen.manila:1995,matousek.thomas:1992,nishimura.ea:2000}.
Efficient algorithms are known for subtree
isomorphism~\cite{shamir.tsur:1999,valiente:2002}, for subtree
homeomorphism~\cite{chung:1987,valiente:cpm.2003,valiente:cpm:jda},
for largest common subtree under isomorphic
embeddings~\cite{valiente:2002} and homeomorphic
embeddings~\cite{pinter.ea:2004}, and for both largest common subtree
and smallest common supertree under isomorphic and topological
embeddings~\cite{gupta.nishimura:1998}.  The only (exponential)
algorithm known for largest common subtree under minor embeddings is
given in \cite{shasha.wang.ea:1994}.

Particular cases of these embedding problems for trees have also been
thoroughly studied in the literature.  On ordered trees, they become
polynomial-time solvable for isomorphic, homeomorphic, topological,
and also minor embeddings.  In this particular case, the largest
common subtree problem under homeomorphic embeddings is known as the
maximum agreement subtree
problem~\cite{amir.keselman:1997,cole.ea:2000,steel.warnow:1993}, the
largest common subtree problem under minor embeddings is known as the
tree edit
problem~\cite{dulucq.tichit:2003,shasha.zhang:1990,zhang.shasha:1989},
and the smallest common supertree problem under minor embeddings is
known as the tree alignment
problem~\cite{jansson.lingas:2003,jiang.ea:1995,wang.zhao:2003}.  The
smallest common supertree problem under minor embeddings was also
studied in~\cite{nishimura.ea:2000} for trees of bounded degree.

In this paper, we establish in a unified way the relationship between
the largest common subtree and the smallest common supertree problems
for isomorphic, homeomorphic, topological, and minor embeddings.  A
similar correspondence between largest common subgraphs and smallest
common supergraphs under isomorphic embeddings was studied
in~\cite{fernandez.valiente:2001}.  More specifically, we give a
simple and unique construction that allows one to obtain in all four
cases a largest common subtree of two trees from any smallest common
supertree of them, and vice versa, another simple and unique
construction that allows one to obtain in all four cases a smallest
common supertree of two trees from any largest common subtree of them.
These constructions take only time linear in the size of the trees,
and, moreover, they have a clear algebraic meaning: in all four types
of embeddings, a largest common subtree of two trees is obtained as
the \emph{pullback} of their embeddings into a smallest common
supertree, and a smallest common supertree of two trees is obtained as
the \emph{pushout} of the embeddings of a largest common subtree into
them.  This is, to the best of our knowledge, the first unified
construction showing the relation between largest common subtrees and
smallest common supertrees for isomorphic, homeomorphic, topological,
and minor embeddings.  These results answer the open problem of
establishing the relationship between the largest common subtree and
the smallest common supertree under any embedding relation, posed by
the last author in his talk ``Subgraph Isomorphism and Related
Problems for Restricted Graph Classes'' at Dagstuhl Seminar 04221,
``Robust and Approximative Algorithms on Particular Graph Classes,''
May 23--28, 2004.

Roughly speaking, our constructions work as follows.  Given two trees
$T_1$ and $T_2$ and a largest common subtree $T_{\mu}$ explicitly
embedded into them, a smallest common supertree of $T_1$ and $T_2$ is
obtained by first making the disjoint sum of $T_{1}$ and $T_{2}$, then
merging in this sum each two nodes of $T_{1}$ and $T_{2}$ that are
related to the same node of $T_{\mu}$, and finally removing all
parallel arcs and all arcs subsumed by paths.  Conversely, given two
trees $T_1$ and $T_2$ embedded into a smallest common supertree $T$ of
them, a largest common subtree of $T_1$ and $T_2$ is obtained by
removing all nodes in $T$ not coming from both $T_1$ and $T_2$, and
then replacing by arcs all paths between pairs of remaining nodes that
do not contain other remaining nodes.  Unfortunately, the
justification for these simple constructions, as well as the proof of
their algebraic meaning, is rather intricate, and at some points it
differs substantially for the different notions of embedding.

Beyond their theoretical interest, these constructions provide an
efficient solution of the smallest common supertree problem under
homeomorphic embeddings, for which no algorithm was known until now.
The solution extends the largest common homeomorphic subtree algorithm
of~\cite{pinter.ea:2004}, which in turn extended the subtree
homeomorphism algorithm of~\cite{valiente:cpm.2003,valiente:cpm:jda}.
Likewise, these constructions also provide a solution to the
smallest common supertree problem under minor embeddings, for which no
algorithm was known previously, either.  The solution extends the
unordered tree edit algorithm of~\cite{shasha.wang.ea:1994}.

\section{Preliminaries} \label{sec-prel}

In this section we recall the categorical notions of pushouts and
pullbacks, as they are needed in the following sections, and the
notions of isomorphic, homeomorphic, topological, and minor embeddings
of trees, together with some results about them that will be used in
the rest of the paper.

\subsection{Pushouts and pullbacks}
A \emph{category} is a structure consisting of: a class of
\emph{objects}; for every pair of objects $A,B$, a class $\Mor(A,B)$
of \emph{morphisms}; and, for every objects $A,B,C$, a binary
operation
$$
\begin{array}{rrcl}
\circ & :\Mor(A,B)\times \Mor(B,C) & \to &  \Mor(A,C)\\
& (f,g) & \mapsto & g\circ f
\end{array}
$$
called \emph{composition}, which satisfies the following two
properties:
\begin{description}
\item[{Associativity}:] for every $f\in \Mor(A,B)$, $g\in
\Mor(B,C)$, and $h\in \Mor (C,D)$, $h\circ(g\circ f)=(h\circ g)\circ
f\in \Mor(A,D)$.

\item[{Existence of identities}:] for every object $A$, there exists an
\emph{identity morphism} $\Id_{A}\in \Mor(A,\allowbreak A)$ such that
$\Id_{A}\circ f=f$, for every $f\in \Mor (B,A)$, and $g\circ
\Id_{A}=g$,
for every $g\in \Mor(A,B)$.
\end{description}
It is usual to indicate that $f\in \Mor(A,B)$ by writing $f:A\to B$.

All categories considered in this paper have all
trees as objects and different types of embeddings of trees as
morphisms: see the next subsection.

A \emph{pushout} in a category $\mathcal{C}$ of two morphisms  
$f_{1}:A\to B_{1}$
and $f_{2}:A\to B_{2}$ is an object $P$ together with two morphisms
$g_{1}:B_{1}\to P$ and $g_{2}:B_{2}\to P$ satisfying the following two
conditions:
\begin{itemize}
\item[(i)] $g_1 \circ f_1=g_2 \circ f_2$.

\item[(ii)] (\emph{Universal property}) If $X$ is any object together
with a pair of morphisms $g'_{1}:B_{1}\to X$ and $g'_{2}:B_{2}\to X$
such that $g'_1 \circ f_1=g'_2 \circ f_2$, then there exists a unique
morphism $h:P \to X$ such that $h \circ g_1=g'_1$ and $h \circ
g_2=g'_2$.
\end{itemize}
% $$
% \xymatrix{
% A \ar[r]^{f_1} \ar[d]_{f_2}
%    & B_1 \ar[d]^{g_1} \ar@/^1pc/[ddr]^{g'_1}  \\
%     B_2 \ar[r]_{g_2} \ar@/_1pc/[drr]_{g'_2}
%    & P \ar@{.>}[dr]|{h} \\
% & & X
% }
% $$
% 

A \emph{pullback} in  a category $\mathcal{C}$ of two morphisms  
$f_{1}:A_{1}\to B$
and $f_{2}:A_{2}\to B$ is an object $Q$ together with two morphisms
$g_1:Q \to A_1$ and $g_2:Q \to A_2$ satisfying the following two
conditions:
\begin{itemize}
\item[(i)] $f_1 \circ g_1=f_2 \circ g_2$.

\item[(ii)] (\emph{Universal property})
If $X$ is any object together  with a pair of morphisms
$g'_1:X \to A_1$ and $g'_2:X \to A_2$ such that $f_1 \circ g'_1=f_2
\circ g'_2$, then there exists a unique morphism $h:X \to Q$ such that
$g'_1=g_1 \circ h$ and $g'_2=g_2 \circ h$.
\end{itemize}
% $$
% \xymatrix{
%   X \ar@/^1pc/[drr]^{g'_1} \ar@/_1pc/[ddr]_{g'_2} \ar@{.>}[dr]|{h} \\
%   & Q \ar[r]^{g_1} \ar[d]_{g_2}
%    & A_1 \ar[d]^{f_1}\\
%    & A_2 \ar[r]_{f_2}
%    & B
%   }
% $$
% 

Two pushouts in $\mathcal{C}$ of the same pair of morphisms, as well as
two pullbacks in $\mathcal{C}$ of the same pair of morphisms, are
always isomorphic in $\mathcal{C}$.

% A category $\mathcal{C}$ \emph{has all binary pushouts} if every
% pair of morphisms with the same source object has a pushout, and it
% \emph{has all binary pullbacks} if every pair of morphisms with the
% same target object has a pullback.

\subsection{Embeddings of trees}

A \emph{directed graph} is a structure $G=(V,E)$ consisting of a set
$V$, whose elements are called \emph{nodes}, and a set $E$ of ordered
pairs $(a,b)\in V\times V$ with $a\neq b$; the elements of $E$ are
called \emph{arcs}.  For every arc $(v,w)\in E$, $v$ is its
\emph{source} node and $w$ its \emph{target} node.  A graph is
\emph{finite} if its set of nodes is finite.  The \emph{in-degree} of
a node $v$ in a finite graph is the number of arcs that have $v$ as
target node and its \emph{out-degree} is the number of arcs that
have $v$ as source node.

An \emph{isomorphism} $f:G\to G'$ between graphs $G=(V,E)$ and $G'=(V',E')$
is a bijective mapping $f:V\to V'$ such that, for
every $a,b\in V$, $(a,b)\in E$ if and only if $(f(a),f(b))\in E'$.

A \emph{path} in a directed graph $G=(V,E)$ is a sequence of nodes
$(v_{0},v_{1},\ldots,v_{k})$ such that $(v_{0},v_{1}),(v_{1},v_{2}),
(v_{2},v_{3}),\ldots,(v_{k-1},v_{k})\in E$; its \emph{origin} is
$v_{0}$, its \emph{end} is $v_{k}$, and its \emph{intermediate nodes}
are $v_{1},\ldots,v_{k-1}$.  Such a path is \emph{non-trivial} if
$k\geq 1$.  We shall represent a path \emph{from $a$ to $b$}, that is,
a path with origin $a$ and end $b$, by $a\cami b$.

A (\emph{rooted}) \emph{tree} is a directed finite graph $T=(V,E)$
with $V$ either empty or containing a distinguished node $r\in V$,
called the \emph{root}, such that for every other node $v\in V$ there
exists one, and only one, path $r\cami v$.  Note that every node in a
tree has in-degree 1, except the root that has in-degree 0.
Henceforth, and unless otherwise stated, given a tree $T$ we shall
denote its set of nodes by $V(T)$ and its set of arcs by $E(T)$.  The
\emph{size} of a tree $T$ is its number $|E(T)|$ of arcs.

The \emph{children} of a node $v$ in a tree $T$ are those nodes $w$
such that $(v,w)\in E(T)$: in this case we also say that $v$ is the
\emph{parent} of its children.  The only node without  parent is the
root, and the nodes without children are the \emph{leaves} of the
tree.

A path $(v_{0},v_{1},\ldots,v_{k})$ in a tree $T$ is \emph{elementary}
if, for every $i=1,\ldots,k-1$, $v_{i+1}$ is the only child of
$v_{i}$; in other words, if all its intermediate nodes have
out-degree 1.  In particular, an arc forms an elementary path.

Two non-trivial paths $(a,v_{1},\ldots,v_{k})$ and
$(a,w_{1},\ldots,w_{\ell})$ in a tree $T$ are said to \emph{diverge}
if their origin $a$ is their only common node.  Note that, by the
uniqueness of paths in trees, this condition is equivalent to
$v_{1}\neq w_{1}$.  The definition of trees also implies that, for
every two nodes $b,c$ of a tree that are not connected by a path,
there exists one, and only one, node $a$ such that there exist
divergent paths $a\cami b$ and $a\cami c$: we shall call this
node the \emph{least common ancestor} of $b$ and $c$.  The adjective
``least'' refers to the obvious fact that if there exist paths from a
node $x$ to $b$ and to $c$, then these paths consist of a path from
$x$ to the least common ancestor of $b$ and $c$ followed by the
divergent paths from this node to $b$ and $c$.

\begin{definition}
Let $S$ and $T$ be trees.
\begin{enumerate}
\renewcommand{\labelenumi}{(\roman{enumi})}

%%minor
\item  $S$ is a \emph{minor} of  $T$ if there exists an
injective mapping $f:V(S)\to V(T)$ satisfying the following condition:
for every $a,b\in V(S)$, if  $(a,b)\in E(S)$,
then there exists a path  $f(a)\cami f(b)$ in $T$ with no
intermediate node in $f(V(S))$.  In this case, the mapping $f$ is said
to be a \emph{minor embedding} $f:S \to T$.

%%top
\item $S$ is a \emph{topological subtree} of $T$ if there exists a
minor embedding $f:S\to T$ such that, for every $(a,b),(a,c)\in E(S)$
with $b\neq c$, the paths $f(a)\cami f(b)$ and $f(a)\cami f(c)$ in $T$
diverge.  In this case, $f$ is called a \emph{topological embedding}
$f:S \to T$.

%%hom
\item  $S$ is a \emph{homeomorphic subtree} of  $T$ if there
exists a minor embedding $f:S\to T$ satisfying the following extra
condition: for every $(a,b)\in E(S)$, the path  $f(a)\cami f(b)$
in $T$ is elementary.  In this case, $f$ is said to be a
\emph{homeomorphic embedding} $f:S \to T$.

%%iso
\item  $S$ is an \emph{isomorphic subtree} of  $T$ if there
exists an injective mapping $f:V(S)\to V(T)$ satisfying the following
condition: if $(a,b)\in E(S)$, then $(f(a),f(b))\in E(T)$.  Such a
mapping $f$ is called an \emph{isomorphic embedding} $f:S \to T$.
\end{enumerate}
\end{definition}

\begin{lemma}\label{implicacions}
Every isomorphic embedding is a homeomorphic embedding, every
homeomorphic embedding is a topological embedding, and every
topological embedding is a minor embedding.
\end{lemma}

\begin{proof}
It is obvious from the definitions that every isomorphic embedding is
a homeomorphic embedding and that every topological embedding is a
minor embedding.  Now, let $f:S\to T$ be a homeomorphic embedding and
let $(a,b),(a,c)\in E(S)$ be such that $b\neq c$.  Then, the paths
$f(a)\cami f(b)$ and $f(a)\cami f(c)$ are elementary and they do not
contain any intermediate node in $f(V(S))$.  This implies that neither
$f(b)$ is intermediate in the path $f(a)\cami f(c)$, nor $f(c)$ is
intermediate in the path $f(a)\cami f(b)$. Therefore, $f(b)$ and
$f(c)$ are not connected by a path.  But then the least common
ancestor $x$ of $f(b)$ and $f(c)$ must have out-degree at least 2, and
thus it cannot be intermediate in the paths from $f(a)$ to these
nodes.  Since there exists a path $f(a)\cami x$, we conclude that
$f(a)=x$, that is, the paths $f(a)\cami f(b)$ and $f(a)\cami f(c)$
diverge.  This shows that $f$ is a topological embedding.  \qed
\end{proof}

The implications in the last lemma are strict, as the following
example shows.

\begin{figure}[htb]
\begin{center}
     \begin{picture}(200,90)(0,0)
      \setlength{\unitlength}{1.2pt}
\put(0,0){\line(1,1){30}}
\put(60,0){\line(-1,1){30}}
\put(0,0){\circle*{3}}
\put(60,0){\circle*{3}}
\put(30,30){\circle*{3}}
\put(-5,0){\makebox(0,0)[r]{$x$}}
\put(65,0){\makebox(0,0)[l]{$y$}}
\put(35,30){\makebox(0,0)[l]{$r$}}
\put(60,30){\makebox(0,0)[b]{$S$}}
%%%
\put(120,40){\line(1,1){20}}
\put(160,0){\line(1,1){20}}
\put(200,0){\line(-1,1){60}}
\put(120,40){\circle*{3}}
\put(140,60){\circle*{3}}
\put(160,40){\circle*{3}}
\put(180,20){\circle*{3}}
\put(200,0){\circle*{3}}
\put(160,0){\circle*{3}}
\put(115,40){\makebox(0,0)[r]{$2$}}
\put(155,0){\makebox(0,0)[r]{$5$}}
\put(205,0){\makebox(0,0)[l]{$6$}}
\put(185,20){\makebox(0,0)[l]{$4$}}
\put(165,40){\makebox(0,0)[l]{$3$}}
\put(145,60){\makebox(0,0)[b]{$1$}}
\put(190,50){\makebox(0,0)[b]{$T$}}
\end{picture}
\end{center}
\caption{\label{fig:sec2.2.1}%
The trees $S$ and $T$ in Example \ref{ex:fig:sec2.2.1}}
\end{figure}
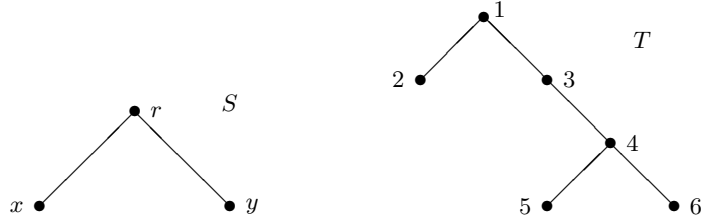

\begin{example}\label{ex:fig:sec2.2.1}
Let $S$ and $T$ be the trees described in
Fig.~\ref{fig:sec2.2.1}, with roots $r$ and $1$, respectively.

\begin{enumerate}
\item[(a)] The mapping $f_{0}:V(S)\to V(T)$ defined by $f_{0}(r)=1$,
$f_{0}(x)=3$ and $f_{0}(y)=4$ is not a minor embedding, because,
although it transforms arcs in $S$ into paths in $T$, the path
$f_{0}(r)\cami f_{0}(y)$ contains the node $3=f_{0}(x)$, which
belongs to $f_{0}(V(S))$.

\item[(b)] The mapping $f_{1}:V(S)\to V(T)$ defined by $f_{1}(r)=1$,
$f_{1}(x)=5$ and $f_{1}(y)=6$ is a minor embedding, because the arcs
$(r,x),(r,y)\in E(S)$ become paths $f_{1}(r)\cami f_{1}(x)$ and
$f_{1}(r)\cami f_{1}(y)$ in $T$ with no intermediate node in
$f_{1}(V(S))$.  But it is not a topological embedding, because these
paths do not diverge.

\item[(c)] The mapping $f_{2}:V(S)\to V(T)$ defined by $f_{2}(r)=1$,
$f_{2}(x)=2$ and $f_{2}(y)=6$ is a topological embedding, because the
arcs $(r,x),(r,y)\in E(S)$ become divergent paths $f_{2}(r)\cami
f_{2}(x)$ and $f_{2}(r)\cami f_{2}(y)$ in $T$ without intermediate
nodes in $f_{2}(V(S))$.  But it is not a homeomorphic embedding,
because the path $f_{2}(r)\cami f_{2}(y)$ contains an intermediate
node with more than one child.

\item[(d)] The mapping $f_{3}:V(S)\to V(T)$ defined by $f_{3}(r)=1$,
$f_{3}(x)=2$ and $f_{3}(y)=4$ is a homeomorphic embedding, because the
arcs $(r,x),(r,y)\in E(S)$ become elementary paths  $f_{3}(r)\cami
f_{3}(x)$ and $f_{3}(r)\cami f_{3}(y)$ in $T$ with no intermediate node  
in
$f_{3}(V(S))$.  But it is not an isomorphic embedding, because the
path  $f_{3}(r)\cami f_{3}(y)$ is not an arc.

\item[(e)] The mappings $f_{4}:V(S)\to V(T)$ defined by $f_{4}(r)=1$,
$f_{4}(x)=2$ and $f_{4}(y)=3$, and $f_{5}:V(S)\to V(T)$ defined by
$f_{5}(r)=4$, $f_{5}(x)=5$ and $f_{5}(y)=6$ are isomorphic embeddings,  
because
they transform every arc in $S$ into an arc in $T$.
\end{enumerate}
\end{example}

The following lemmas will be used several times in the next sections.

\begin{lemma} \label{lem-path}
Let $f:S\to T$ be a minor embedding.  For every $a,b\in V(S)$, there
exists a path $a\cami b$ in $S$ if and only if there exists a path
$f(a)\cami f(b)$ in $T$.  Moreover, if the path $f(a)\cami f(b)$ is
elementary, then the path $a\cami b$ is also elementary, and if there
is an arc from $f(a)$ to $f(b)$ in $T$, then there is an arc from $a$
to $b$ in $S$.
\end{lemma}

\begin{proof}
Since the arcs in $S$ become paths in $T$ without intermediate nodes
in $f(V(S))$, it is obvious that a path $a\cami b$ in $S$ becomes,
under $f$, a path $f(a)\cami f(b)$ in $T$ whose intermediate nodes
belonging to $f(V(S))$ are exactly the images under $f$ of the
intermediate nodes of the path $a\cami b$.

Assume now that there exists a path $f(a)\cami f(b)$ in $T$, and let
$r$ be the root of $S$.  If $a=r$ or $a=b$, it is clear that there
exists a path $a\cami b$ in $S$.  If $a\neq r$ and $a\neq b$, then the
images of the paths $r\cami a$ and $r\cami b$ in $S$ are paths
$f(r)\cami f(a)$ and $f(r)\cami f(b)$ in $T$.  Now, the uniqueness of
paths in $T$ implies that the path $f(r)\cami f(b)$ splits into the
path $f(r)\cami f(a)$ and the path $f(a)\cami f(b)$. Therefore,
$f(a)$ is an intermediate node of the path $f(r)\cami f(b)$.
As a consequence, since $f$ is injective and any intermediate node of this path
belonging to $f(V(S))$ must be the image under $f$ of an intermediate
node of the path $r\cami b$, the node $a$ must  be intermediate
in the path $r\cami b$, which yields a path $a\cami b$ in
$S$.

Moreover, if a node in $S$ has more than one child, then its image
under $f$ has also more than one child.  This implies that if the path
$f(a)\cami f(b)$ is elementary, then the path $a\cami b$ is
elementary, too.  Finally, if there is an arc from $f(a)$ to $f(b)$,
then the path $a\cami b$ cannot have any intermediate node: it must be
an arc.  \qed
\end{proof}

By Lemma \ref{implicacions}, the last lemma applies also to
isomorphic, homeomorphic, and topological embeddings.

\begin{lemma} \label{lem-lca}
Let $f:S\to T$ be a topological embedding.  For every $a,b\in V(S)$ not
connected by a path, if $x$ is their least common ancestor in $S$, then
$f(x)$ is the least common ancestor of $f(a)$ and $f(b)$ in $T$.
\end{lemma}

\begin{proof}
Since $a$ and $b$ are not connected by a path in $S$, by the last
lemma we know that $f(a)$ and $f(b)$ are not connected by a path in
$T$, either.  Let now $x$ be the least common ancestor of $a$ and $b$
in $S$, and let $v$ and $w$ be the children of $x$ contained in the
divergent paths  $x\cami a$ and $x\cami b$, respectively.  Then, since  
$f$ is a
topological embedding, there exist in $T$ divergent paths  $f(x)\cami
f(v)$ and  $f(x)\cami f(w)$, which are followed by paths  $f(v)\cami
f(a)$ and  $f(w)\cami f(b)$, respectively.  This means that
$f(x)$ is the node in $T$ from which there exist divergent paths to
$f(a)$ and to $f(b)$, that is, the least common ancestor of these two
nodes.\qed
\end{proof}

By Lemma \ref{implicacions}, the last lemma applies also to isomorphic
and homeomorphic embeddings.  But the thesis of this lemma need not
hold if $f$ is only a minor embedding: see, for instance
Example \ref{ex:fig:sec2.2.1}.(b), where $r$ is the least common
ancestor of $x$ and $y$, but the least common ancestor of $f_{1}(x)=5$
and $f_{1}(y)=6$ is $4$, and not $f_{1}(r)=1$.

\begin{lemma}\label{lem-iso}
Every bijective minor embedding is an isomorphism of graphs.
\end{lemma}

\begin{proof}
Let $f:S\to T$ be a minor embedding such that $f:V(S)\to 
V(T)$ is bijective, and let $a,b\in V(S)$. 
If $(a,b)\in E(S)$, then there exists a path 
$f(a)\cami f(b)$ in $T$ without any intermediate node in $f(V(S))$. 
Since $f$ is bijective, this means that this path has no intermediate 
node, and thus it is an arc. This proves that if $(a,b)\in E(S)$, 
then $(f(a),f(b))\in E(T)$. The converse implication is given by
Lemma \ref{lem-path}.\qed
\end{proof}

By Lemma \ref{implicacions}, the last lemma implies that every
bijective isomorphic, homeomorphic, or topological embedding is an
isomorphism of graphs.

\begin{definition}
Let $S$ and $T$ be trees.
\begin{enumerate}
     \renewcommand{\labelenumi}{(\roman{enumi})}

\item A \emph{largest common isomorphic subtree (homeomorphic subtree,
topological subtree, minor)} of  $S$ and $T$ is a tree that
is an isomorphic subtree (respectively, homeomorphic subtree,  
topological subtree,
minor) of both of them and has the largest size among all trees with
this property.

\item A \emph{smallest common isomorphic supertree (homeomorphic  
supertree,
top\-ological supertree, supertree under minor embeddings)} of
$S$ and $T$ is a tree such that both $S$ and $T$ are isomorphic
subtrees (respectively, homeomorphic subtrees, topological subtrees,  
minors) of it
and has the least size among all trees with this property.
\end{enumerate}
\end{definition}

We shall denote by $\CC_{iso}$, $\CC_{hom}$, $\CC_{top}$, and
$\CC_{min}$ the categories with objects all trees and with morphisms
the isomorphic, homeomorphic, topological, and minor embeddings,
respectively.  Whenever we denote generically any one of these
categories by $\CC_{*}$, we shall use the following notations.  By a
\emph{$\CC_{*}$-embedding} we shall mean a morphism in the
corresponding category.  By a \emph{common $\CC_{*}$-subtree} of two
trees we shall mean a tree together with $\CC_{*}$-embeddings into
these two trees.  By a \emph{largest common $\CC_{*}$-subtree} of two
trees we shall mean a largest size common $\CC_{*}$-tree.  By a
\emph{common $\CC_{*}$-supertree} of two trees we shall mean a
tree together with $\CC_{*}$-embeddings of these two trees into it.
By a \emph{smallest common $\CC_{*}$-supertree} of two trees we
shall mean a least size common $\CC_{*}$-supertree.
And by a \emph{$\CC_{*}$-path} we shall
understand an arc if $\CC_{*}$ stands for $\CC_{iso}$, an elementary
path if $\CC_{*}$ denotes $\CC_{hom}$, and an arbitrary path if
$\CC_{*}$ means $\CC_{top}$ or $\CC_{min}$.  Note in particular that
all trivial paths and all arcs are $\CC_{*}$-paths, for every category
$\CC_{*}$.

The following corollary is a simple rewriting of the definitions.

\begin{corollary}\label{corsec2.1}
Let $\CC_{*}$ denote any category $\CC_{iso}$, $\CC_{hom}$, or
$\CC_{min}$.  For every trees $S,T$, a mapping $f:V(S)\to V(T)$ is a
$\CC_{*}$-embedding if and only if, for every $(a,b)\in E(S)$, there
is a $\CC_{*}$-path $f(a)\cami f(b)$ in $T$ with no intermediate node
belonging to $f(V(S))$.
\end{corollary}

And the following corollary is a direct consequence of  
Lemma~\ref{lem-path}.

\begin{corollary} \label{lem-path-2}
Let $\CC_{*}$ be any category $\CC_{iso}$, $\CC_{hom}$,
$\CC_{top}$, or $\CC_{min}$, and let $f:S\to T$ be a
$\CC_{*}$-embedding.  For every $a,b\in V(S)$, if there exists a
$\CC_{*}$-path  $f(a)\cami f(b)$ in $T$, then there exists a
$\CC_{*}$-path  $a\cami b$ in $S$.
\end{corollary}

Finally, we have the following result, which will be used later.

\begin{lemma}\label{seg-inic}
Let $\CC_{*}$ be any category $\CC_{iso}$, $\CC_{hom}$, $\CC_{top}$,
or $\CC_{min}$, let $S,T,U$ be trees and $f:V(S)\to V(T)$ and
$g:V(T)\to V(U)$ mappings between their sets of nodes.  If $g\circ
f:S\to U$ and $g:T\to U$ are $\CC_{*}$-embeddings, then $f:S\to T$ is
also a $\CC_{*}$-embedding.
\end{lemma}

\begin{proof}
Since $g\circ f$ is injective, it is clear that $f$ is injective.  Let
now $a,b\in S$ be such that $(a,b)\in E(S)$.  Since $g\circ f:S\to U$
is a $\CC_{*}$-embedding, there exists a $\CC_{*}$-path $g(f(a))\cami
g(f(b))$ in $U$ without any intermediate node in $g(f(V(S)))$.  Since
$g:T\to U$ is a $\CC_{*}$-embedding, the existence of this path
$g(f(a))\cami g(f(b))$ in $U$ implies, by Corollary~\ref{lem-path-2},
the existence of a $\CC_{*}$-path $f(a)\cami f(b)$ in $T$.  This path
cannot have any intermediate node in $f(V(S))$, because any such
intermediate node would become, under $g$, an intermediate node
belonging to $g(f(V(S)))$ in the path $g(f(a))\cami g(f(b))$.  

So, $f$ is injective and if $(a,b)\in E(S)$, then there exists a
$\CC_{*}$-path $f(a)\cami f(b)$ in $T$ without intermediate nodes in
$f(V(S))$.  This already shows, by Corollary \ref{corsec2.1}, that $f$
is a $\CC_{*}$-embedding when $\CC_{*}$ stands for $\CC_{iso}$,
$\CC_{hom}$, or $\CC_{min}$.

As far as $\CC_{top}$ goes, we have already proved that $f$ transforms
arcs into paths without intermediate nodes in $f(V(S))$, and thus it
remains to prove that if $a,b,c\in V(S)$ are such that $(a,b),(a,c)\in
E(S)$ and $b\neq c$, then the paths $f(a)\cami f(b)$ and $f(a)\cami
f(c)$ in $T$ diverge.  But since $g\circ f$ is a topological
embedding, the paths $g(f(a))\cami g(f(b))$ and $g(f(a))\cami g(f(c))$
in $U$ are divergent, and this clearly implies that the paths
$f(a)\cami f(b)$ and $f(a)\cami f(c)$ in $T$ are divergent, too: any
common intermediate node in these paths would become, under $g$, a
common intermediate node in the paths $g(f(a))\cami g(f(b))$ and
$g(f(a))\cami g(f(c))$.  \qed
\end{proof}

%%%%%%%

\section{Common subtrees as pullbacks} \label{sec-pb}

In this section we study the construction of common subtrees as
pullbacks of embeddings into common supertrees, for each one of the
types of tree embeddings considered in this paper.  We start with the
most general type, minor embeddings.

Let $f_{1}:T_{1}\to T$ and $f_{2}:T_{2}\to T$ be henceforth two minor
embeddings.  Without any loss of generality, and unless otherwise
stated, we shall assume that $V(T_{1}),V(T_{2})\subseteq V(T)$ and
that the minor embeddings $f_{1}$ and $f_{2}$ are given by these
inclusions.  For simplicity, we shall denote thus the image of a node
$a\in V(T_{i})$ under the corresponding $f_{i}$ again by $a$.

Let $T_{p}$ be the graph with set of nodes $V(T_{p})=V(T_{1})\cap
V(T_{2})$ and set of arcs defined in the following way: for every
$a,b\in V(T_{1})\cap V(T_{2})$, $(a, b)\in E(T_{p})$ if and only if  
there
are paths $a\cami b$ in $T_{1}$ and in $T_{2}$ without
intermediate nodes in $V(T_{1})\cap V(T_{2})$.
We shall call this graph $T_{p}$ the \emph{intersection} of $T_{1}$
and $T_{2}$ obtained through $f_{1}$ and $f_{2}$.

This graph satisfies the following useful lemma.

\begin{lemma} \label{lem-tpo}
For every $a,b\in V(T_{1})\cap V(T_{2})$:
\begin{enumerate}
\renewcommand{\labelenumi}{(\roman{enumi})}
\item If there exists a
path $a\cami b$ in $T_{p}$, then there exist paths $a\cami b$ in $T_{1}$
and in $T_{2}$.

\item If there exists a path $a\cami b$ in some $T_{i}$, $i=1,2$, then
there exists also a path $a\cami b$ in $T_{p}$, and its intermediate
nodes are exactly the intermediate nodes of the path $a\cami b$ in
$T_{i}$ that belong to $V(T_{1})\cap V(T_{2})$.
\end{enumerate}
\end{lemma}

\begin{proof}
Point (i) is a direct consequence of the fact that every arc in
$T_{p}$ corresponds to paths in $T_{1}$ and $T_{2}$.

As far as point (ii) goes, we shall prove that if there exists a path
$a\cami b$ in $T_{1}$, then there exists also a path $a\cami b$ in
$T_{p}$ with intermediate nodes the intermediate nodes of the path in
$T_{1}$ that belong to $V(T_{1})\cap V(T_{2})$, by induction on the
number $n$ of such intermediate nodes belonging to $V(T_{1})\cap
V(T_{2})$.

If $n=0$, then there exists a path $a\cami b$ in $T_{1}$ that does not
contain any intermediate node in $V(T_{1})\cap V(T_{2})$.  Since
$f_{1}$ transforms arcs into paths with no intermediate node belonging
to $T_{1}$, this implies that there exists a path $a\cami b$ in $T$
that does not contain any node in $V(T_{1})\cap V(T_{2})$, either.
Then, by Lemma~\ref{lem-path}, this path is induced by a path $a\cami
b$ in $T_{2}$, and by the same reason this path does not contain any
intermediate node in $V(T_{1})\cap V(T_{2})$.  So, there are paths
$a\cami b$ in $T_{1}$ and $T_{2}$ without  intermediate
nodes in $V(T_{1})\cap V(T_{2})$, and therefore, by definition, there
exists an arc from $a$ to $b$ in $T_{p}$.

As the induction hypothesis, assume that the claim is true for paths
in $T_{1}$ with $n$ intermediate nodes in $V(T_{1})\cap V(T_{2})$, and
assume now that the path $a\cami b$ has $n+1$ such nodes.  Let $a_{0}$
be the first intermediate node of this path belonging to $V(T_{1})\cap
V(T_{2})$.  Then, by the case $n=0$, there is an arc in $T_{p}$ from
$a$ to $a_{0}$, and by the induction hypothesis there is a path
$a_{0}\cami b$ in $T_{p}$ whose only intermediate nodes are the
intermediate nodes of the path $a_{0}\cami b$ in $T_{1}$  that belong
to $V(T_{1})\cap V(T_{2})$; by concatenating these paths in $T_{p}$ we  
obtain the path
$a\cami b$ we were looking for.  \qed
\end{proof}

The intersection of two minors need not be a tree, as the following
simple example shows.

\begin{example} \label{ex-min}
Let $T$ be a tree with nodes $a_{1},a_{2},b,c$ and arcs
$(a_{1},a_{2}),(a_{2},b),\allowbreak (a_{2},c)$, let $T_{1}$ be its
minor with nodes $a_{1},b,c$ and arcs
$(a_{1},b),(a_{1},c)$, and let $T_{2}$ be its minor with 
nodes $a_{2},b,c$ and arcs $(a_{2},b),(a_{2},c)$.  In this
case $T_{p}$ is the graph with nodes $b,c$ and no arc, and in
particular it is not a tree.
\end{example}

Now  we have the following result.

\begin{proposition} \label{cond-I-min}
$T_{1}$ and $T_{2}$ have always a common minor, which is either
$T_{p}$ together with its inclusions in $T_{1}$ and $T_{2}$,
or  obtained by adding a root to $T_{p}$.
\end{proposition}

\begin{proof}
If $T_{p}$ is empty, then it is a tree and its inclusions into $T_{1}$
and $T_{2}$ are clearly minor embeddings.  In this case, $T_{p}$ is a
common minor of $T_{1}$ and $T_{2}$.

So, assume in the sequel that $T_{p}$ is non-empty.  If it had no node
without parents, then it would contain a circuit and this would imply,
by Lemma \ref{lem-tpo}.(i), the existence of circuits in the trees
$T_{1}$ and $T_{2}$, which is impossible.  Therefore, $T_{p}$ contains
nodes without parent.  Now we must consider two cases:
\begin{enumerate}
\renewcommand{\labelenumi}{(\arabic{enumi})}
\item $T_{p}$ has only one node $r_{p}$ without a parent.  Then every
other node $a$ in $T_{p}$ can be reached from $r_{p}$ through a path,
because this graph does not contain any circuit (as we have seen) and
hence it must contain a path from a node of in-degree 0 to $a$.  To
check that this path is unique, we shall prove that no node in $T_{p}$
has in-degree greater than 1.

Indeed, assume that there are nodes $a,b,c\in V(T_{p})$, with $b\neq
c$, and arcs from $b$ and $c$ to $a$.  This means that there are paths
in $T_{1}$ and in $T_{2}$ from $b$ and $c$ to $a$ that do not contain
any intermediate node in $V(T_{1})\cap V(T_{2})$.  But since, say,
$T_{1}$ is a tree, if there exist paths $b\cami a$ and $c\cami a$
in $T_{1}$, one of the nodes $b$ or $c$ must be intermediate in the  
path from the
other one to $a$, which yields a contradiction.

This proves that, in this case, $T_{p}$ is a tree.  And by definition,  
for
every $a,b\in V(T_{p})$, if $(a,b)\in T_{p}$,
then there are paths  $a\cami b$ in $T_{1}$ and in $T_{2}$ without
any intermediate node in $V(T_{p})$.  Therefore, the
inclusions $\iota_{i}:V(T_{p})\hookrightarrow V(T_{i})$ induce minor  
embeddings
$\iota_{i}:T_{p}\to T_{i}$, for $i=1,2$, and hence $T_{p}$ is a common
minor of $T_{1}$ and $T_{2}$.

\item $T_{p}$ contains more than one node without a parent, say
$x_{1},\ldots,x_{k}$.  The same argument used in (1) shows in this
case that every other node $a\in V(T_{p})$ can be reached from one of
these nodes $x_{i}$ through a path in $T_{p}$, and that no node in
$T_{p}$ has in-degree greater than 1.

Let now $\tilde{T}_{p}$ be the graph obtained by adding to $T_{p}$ one
node $r$ and arcs $(r,x_{i})$, for $i=1,\ldots,k$.  Then, $r$ is the
only node without a parent in $\tilde{T}_{p}$ and every node in it is
reached from $r$ through a unique path. Indeed, each $x_{i}$ is reached
from $r$ through the new arc  $(r,x_{i})$, and then every
other node in $\tilde{T}_{p}$ is reached from $r$ by the path
going from some $x_{i}$ to it in $T_{p}$ preceded by the arc from $r$
to this $x_{i}$. And these paths are unique, because no node in
$\tilde{T}_{p}$ has in-degree greater than 1.  Therefore,
$\tilde{T}_{p}$ is a tree with root $r$.

Now, note that there is no non-trivial path in either $T_{1}$ or
$T_{2}$ from any node belonging to $V(T_{1})\cap V(T_{2})$ to any
$x_{i}$: such a path, by Lemma~\ref{lem-tpo}, would induce a
non-trivial path in $T_{p}$ and therefore the node $x_{i}$ would have
a parent in $T_{p}$.  This implies in particular that neither the root
of $T_{1}$ nor the root of $T_{2}$ belong to $V(T_{1})\cap V(T_{2})$:
since $k\geq 2$, there are non-trivial paths from each one of these
roots to some $x_{i}$.

Consider then the injective mappings $\widetilde{\iota}_{i}:
\tilde{T}_{p}\to T_{i}$, $i=1,2$, defined by the inclusions on
$V(T_{p})$ and sending $r$ to the root of the corresponding $T_{i}$.
It is clear that they are minor embeddings: on the one hand, arguing
as in (1) above, we obtain that the restriction of each
$\widetilde{\iota}_{i}$ to $T_{p}$ sends every arc to a path in
$T_{i}$ without any intermediate node coming from $\tilde{T}_{p}$; on
the other hand, $\widetilde{\iota}_{i}$ sends every arc $(r,x_{\ell})$
to the path in $T_{i}$ going from its root to $x_{\ell}$, which, as we
saw above, does not contain any intermediate node in $V(T_{1})\cap
V(T_{2})$.  Thus, $\tilde{T}_{p}$ is a common minor of $T_{1}$ and
$T_{2}$.  \qed
\end{enumerate}
\end{proof}

If we restrict ourselves from minor embeddings to topological
embeddings, then only the first case in the last proposition can
happen.

\begin{proposition} \label{pb-top}
If $f_{1}:T_{1}\to T$ and $f_{2}:T_{2}\to T$ are topological
embeddings, then $T_{p}$ is a tree and the inclusions
$V(T_{p})\hookrightarrow V(T_{i})$ are topological embeddings
$\iota_{i}:T_{p}\to T_{i}$, for $i=1,2$, and therefore $T_{p}$ is a
common topological subtree of $T_{1}$ and $T_{2}$.
\end{proposition}

\begin{proof}
Let us prove first of all that if $f_{1}$ and $f_{2}$ are not only
minor but topological embeddings, then $T_{p}$ does not have more than
one node without a parent.  Indeed, assume that $a,b\in V(T_{1})\cap
V(T_{2})$ have no parent in $T_{p}$.  Then, neither $T_{1}$ nor
$T_{2}$ contains any non-trivial path from some node in $V(T_{1})\cap
V(T_{2})$ to $a$ or $b$, because, by Lemma~\ref{lem-tpo}, such a path
would imply a non-trivial path in $T_{p}$ finishing in $a$ or $b$ and
then one of these nodes would have a parent in $T_{p}$.  In
particular, there is no path connecting $a$ and $b$ in either $T_{1}$
or $T_{2}$.  For every $i=1,2$, let $x_{i}\in V(T_{i})$ be the least
common ancestor of $a$ and $b$ in $T_{i}$.  By Lemma \ref{lem-lca},
each $x_{i}$ is also the least common ancestor of $a$ and $b$ in $T$.
But then $x_{1}=x_{2}\in V(T_{1})\cap V(T_{2})$ and therefore both $a$  
and $b$ can be reached
from a node in $V(T_{1})\cap V(T_{2})$ through paths in $T_{1}$ and in
$T_{2}$, which yields a contradiction.

So, since every topological embedding is a minor embedding, from the
proof of Proposition~\ref{cond-I-min} we know that the fact that
$T_{p}$ has at most (and hence, exactly) one node without a parent
implies that it is a tree.  Let us prove now that $\iota_{1}$ is a
topological embedding.  By point (1) in the proof of
Proposition~\ref{cond-I-min}, we already know that it is a minor
embedding.  So, it remains to prove that if there are arcs from $a$ to
$b$ and to $c$ in $T_{p}$, then the paths $a\cami b$ and $a\cami c$ in
$T_{1}$ diverge.

To prove it, note that, since, by the definition of $T_{p}$, the paths
$a\cami b$ and $a\cami c$ in $T_{1}$ and in $T_{2}$ have no
intermediate node in $V(T_{1})\cap V(T_{2})$, neither $b$ nor $c$
appears in the path from $a$ to the other one, and therefore there is
no path connecting $b$ and $c$.  Thus, if, for every $i=1,2$,
$x_{i}\in V(T_{i})$ denotes the least common ancestor of $b$ and $c$
in $T_{i}$, then, arguing as before, we deduce that $x_{1}=x_{2}$ and
in particular that this node belongs to $V(T_{1})\cap V(T_{2})$.

Now, the existence of the paths $a\cami b$ and $a\cami c$ in $T_{1}$,
implies that either $x_{1}=a$ or there exists a non-trivial path in
$T_{1}$ from $a$ to $x_{1}$.  But the paths $a\cami b$ and $a\cami c$
in $T_{1}$ do not contain any intermediate node belonging to
$V(T_{1})\cap V(T_{2})$, and therefore it must happen that $a=x_{1}$ and
the paths $a\cami b$ and $a\cami c$ in $T_{1}$ diverge, as we
wanted to prove.  \qed
\end{proof}

%%%

We have similar results if $f_{1}$ and $f_{2}$ are not only
topological, but homeomorphic or isomorphic embeddings.

\begin{proposition} \label{pb-hom}
If $f_{1}:T_{1}\to T$ and $f_{2}:T_{2}\to T$ are homeomorphic
embeddings, then $T_{p}$ is a tree and the inclusions
$V(T_{p})\hookrightarrow V(T_{i})$ are homeomorphic embeddings
$\iota_{i}:T_{p}\to T_{i}$, for $i=1,2$, and therefore $T_{p}$ is a
common homeomorphic subtree of $T_{1}$ and $T_{2}$.
\end{proposition}

\begin{proof}
We already know from Proposition~\ref{pb-top} that $T_{p}$ is a tree
and that the inclusions $\iota_{1}:T_{p}\to T_{1}$ and
$\iota_{2}:T_{p}\to T_{2}$ are topological embeddings.  It remains to
prove that they are not only topological, but homeomorphic embeddings.
We shall do it for $\iota_{1}:T_{p}\to T_{1}$.

Let $a,b\in V(T_{p})$ be such that $(a,b)\in E(T_{p})$.  Then, by
definition, there exists a path $a\cami b$ in $T_{1}$ without any
intermediate node in $V(T_{1})\cap V(T_{2})$.  Assume that this path
has an intermediate node $x$ with more than one child.  The path
$a\cami b$ induces, under the homeomorphic embedding $f_{1}:T_{1}\to
T$, a path $a\cami b$ in $T$ that contains $x$, and this node has also
more than one child in $T$.  Now, by Lemma~\ref{lem-path}, there is
also a path $a\cami b$ in $T_{2}$.  Since every arc in $T_{2}$
becomes, under the homeomorphic embedding $f_{2}:T_{2}\to T$, an
elementary path in $T$, the nodes in the path $a\cami b$ in $T$ that
do not belong to $V(T_{2})$ have only one child.  Therefore, $x\in
V(T_{2})$ and hence $x\in V(T_{1})\cap V(T_{2})$, which contradicts
the fact that the path $a\cami b$ in $T_{1}$ does not contain any
intermediate node in $V(T_{1})\cap V(T_{2})$.  This proves that this
path is elementary, as we wanted.  \qed
\end{proof}

\begin{proposition} \label{pb-iso}
If $f_{1}:T_{1}\to T$ and $f_{2}:T_{2}\to T$ are isomorphic
embeddings, then $T_{p}$ is a tree and the inclusions
$V(T_{p})\hookrightarrow V(T_{i})$ are isomorphic embeddings
$\iota_{i}:T_{p}\to T_{i}$, for $i=1,2$, and therefore, $T_{p}$ is a
common isomorphic subtree of $T_{1}$ and $T_{2}$.
\end{proposition}

\begin{proof}
We already know from Proposition~\ref{pb-hom} that $T_{p}$ is a tree
and that $\iota_{1}:T_{p}\to T_{1}$ and $\iota_{2}:T_{p}\to T_{2}$ are
homeomorphic embeddings, i.e., that if $(a,b)\in E(T_{p})$, then there
are elementary paths $a\cami b$ in $T_{1}$ and in $T_{2}$ without any
intermediate node in $V(T_{1})\cap V(T_{2})$.  We want to prove that
each one of these paths consists of a single arc, i.e., that $a$ is
the parent of $b$ in both trees.

Let $c_{1}$ be the parent of $b$ in $T_{1}$ and $c_{2}$ the parent of
$b$ in $T_{2}$: they exist because there is a path $a\cami b$ in each
tree.  Then, since $T_{1}$ and $T_{2}$ are isomorphic subtrees of $T$,
both $c_{1}$ and $c_{2}$ are parents of $b$ in $T$, and therefore
$c_{1}=c_{2}\in V(T_{1})\cap V(T_{2})$.  So, the parents in $T_{1}$
and in $T_{2}$ of $b$ are the same and they belong to $V(T_{1})\cap
V(T_{2})$.  Since the paths $a\cami b$ in $T_{1}$ and in $T_{2}$ do
not contain any intermediate node in $V(T_{1})\cap V(T_{2})$ and they
must contain $c_{1}$ and $c_{2}$, respectively, this implies that
$a=c_{1}=c_{2}$, as we claimed.\qed
\end{proof}

We have finally the following result, which gives an algebraic content
to the construction of intersections in $\CC_{iso}$, $\CC_{hom}$, and
$\CC_{top}$.

%%%%%

\begin{proposition} \label{pb-gral}
Let $\CC_{*}$ denote any category $\CC_{iso}$, $\CC_{hom}$, or
$\CC_{top}$.  For every pair of $\CC_{*}$-embeddings $f_{1}:T_{1}\to
T$ and $f_{2}:T_{2}\to T$, $$(T_{p}, \iota_{1}:T_{p}\to
T_{1},\iota_{2}:T_{p}\to T_{2})$$ is a pullback of $f_{1}$ and $f_{2}$
in $\CC_{*}$.
\end{proposition}

\begin{proof}
We know from the previous propositions that, in each case,
$T_{p}$ is a tree and $\iota_{1}:T_{p}\to T_{1}$
and $\iota_{2}:T_{p}\to T_{2}$ are $\CC_{*}$-embeddings, and it is
clear that $f_{1}\circ \iota_{1}=f_{2}\circ \iota_{2}$.  Let us check
now the universal property of pullbacks in $\CC_{*}$.

Let $S$ be any tree and let $g_{1}:S\to T_{1}$ and $g_{2}:S\to T_{2}$
be two $\CC_{*}$-embeddings such that $f_{1}\circ g_{1}=f_{2}\circ
g_{2}$.  Then, at the level of nodes, there exists a unique mapping
$g: V(S)\to V(T_{1})\cap V(T_{2})=V(T_{p})$ such that each $g_{i}$ is  
equal to
$g$ followed by the corresponding  inclusion
$\iota_{i}: V(T_{p})\hookrightarrow V(T_{i})$.  And since each
$\iota_{i}:T_{p}\to T_{i}$ and each composition
$g_{i}=\iota_{i}\circ g:S\to T_{i}$ are $\CC_{*}$-embeddings, Lemma
\ref{seg-inic} implies that $g$ is a $\CC_{*}$-embedding from $S$ to
$T_{p}$. This is the unique $\CC_{*}$ embedding that, when composed
with $\iota_{1}$ and $\iota_{2}$, yields $g_{1}$ and $g_{2}$,
respectively.
\qed
\end{proof}

Therefore, the categories $\CC_{iso}$, $\CC_{hom}$, and $\CC_{top}$
have all binary pullbacks. It is not the case with $\CC_{min}$, as
the following simple example shows.

\begin{remark} \label{rem-pb-min}
The minor embeddings $f_{1}:T_{1}\to T$ and $f_{2}:T_{2}\to T$
corresponding to the minors described in Example~\ref{ex-min} do not
have a pullback in $\CC_{min}$.  Indeed, let $P$, together with
$g_{1}:P\to T_{1}$ and $g_{2}:P\to T_{2}$, be a pullback of them in
$\CC_{min}$.  Then, since $f_{1}\circ g_{1}= f_{2}\circ g_{2}:V(P)\to
V(T)$, we have that $g_{1}(V(P))\subseteq \{b,c\}$ and
$g_{2}(V(P))\subseteq \{b,c\}$ and hence, $P$ being a tree and $g_{1}$
and $g_{2}$ being minor embeddings, there are only two possibilities
for $P$:
\begin{itemize}
\item $P$ is empty.  In this case, if we consider a tree $Q$ with one
node $q$ and no arc, and the minor embeddings $h_{1}:Q\to T_{1}$ and
$h_{2}:Q\to T_{2}$ given by $h_{1}(q)=h_{2}(q)=c$, then $f_{1}\circ
h_{1}=f_{2}\circ h_{2}$ but there is no minor embedding $h:Q\to P$
(because $P$ is empty), which contradicts the definition of pullback.

\item $P$ consists of only one node, say $\{x\}$, and no arc, and
$g_{1}$ and $g_{2}$ send $x$ to the same node, $b$ or $c$, in $T_{1}$
and in $T_{2}$.  But then if we consider the same tree $Q$ as before
and the minor embeddings $h_{1}:Q\to T_{1}$ and $h_{2}:Q\to T_{2}$
that send $q$ to the node different from $g_{1}(x)$ and $g_{2}(x)$,
there is again no minor embedding $h:Q\to P$ such that
$h_{1}=g_{1}\circ h$ and $h_{2}=g_{2}\circ h$, which contradicts the
definition of pullback.
\end{itemize}
\end{remark}

Nevertheless, arguing as in the proof of Proposition~\ref{pb-gral} we
obtain the following result.

\begin{proposition} \label{pb-min}
If $f_{1}:T_{1}\to T$ and $f_{2}:T_{2}\to T$ are minor embeddings such
that $T_{p}$ is a tree, then $(T_{p}, \iota_{1}:T_{p}\to
T_{1},\iota_{2}:T_{p}\to T_{2})$ is a pullback of $f_{1}$ and $f_{2}$
in $\CC_{min}$.
\end{proposition}

\begin{proof}
We know from the proof of Proposition \ref{cond-I-min} that if $T_{p}$
is a tree, then $\iota_{1}:T_{p}\to T_{1}$ and $\iota_{2}:T_{p}\to
T_{2}$ are minor embeddings, and it is clear that $f_{1}\circ
\iota_{1}=f_{2}\circ \iota_{2}$.  Then, exactly the same argument used
in Proposition~\ref{pb-gral} shows that, in this case, $(T_{p},
\iota_{1}:T_{p}\to T_{1},\iota_{2}:T_{p}\to T_{2})$ satisfies the
universal property of pushouts in $\CC_{min}$ .  \qed
\end{proof}

%%%

\section{Common supertrees as pushouts} \label{sec-po}

In this section we study the construction of common supertrees as
pushouts of embeddings of largest common subtrees, for each one of
the types of tree embeddings considered in this paper.
Let $\CC_{*}$ be henceforth any one of the categories of trees
$\CC_{iso}$, $\CC_{hom}$, $\CC_{top}$, or $\CC_{min}$.

Let $T_{1}$ and $T_{2}$ be two trees.  Let $T_{\mu}$ be a largest
common $\CC_{*}$-subtree of them, and let $m_{1}:T_{\mu}\to T_{1}$ and
$m_{2}:T_{\mu}\to T_{2}$ be any $\CC_{*}$-embeddings.  Let
$T_{1}+T_{2}$ be the graph obtained as the disjoint sum of the trees
$T_{1}$ and $T_{2}$: that is,
$$
V(T_{1}+T_{2})=V(T_{1})\sqcup V(T_{2}),\quad
E(T_{1}+T_{2})=E(T_{1})\sqcup E(T_{2}).
$$
Let $\theta$ be the equivalence relation on $V(T_{1})\sqcup V(T_{2})$
defined, up to symmetry, by the following condition:
\begin{quote}
$(a,b)\in \theta$
if and only if $a=b$ or there exists some $c\in V(T_{\mu})$ such that
$a=m_{1}(c)$ and $b=m_{2}(c)$.
\end{quote}
We shall denote the equivalence class modulo $\theta$ of an element
$x\in V(T_{1})\sqcup V(T_{2})$ by $[x]$.

Let $T_{po}$ be the quotient graph of $T_{1}+T_{2}$ by this
equivalence:
\begin{itemize}
\item its set of nodes $V(T_{po})$ is the quotient set
$(V(T_{1})\sqcup V(T_{2}))/\theta$, with elements the equivalence
classes of the nodes of $T_{1}$ or $T_{2}$;

\item its arcs are those induced by the arcs in $T_{1}$ or $T_{2}$, in
the sense that $([a],[b]) \allowbreak\in E(T_{po})$ if and only if
there exist $a'\in[a]$, $b'\in [b]$ and some $i=1,2$ such that
$(a',b')\in E(T_{i})$.
\end{itemize}

Note that every equivalence class $[a]\in V(T_{po})$ is either a
2-elements set $\{m_{1}(x),m_{2}(x)\}$, with $x\in V(T_{\mu})$, or a
singleton $\{a\}$, with $a\in V(T_{i})-m_{i}(V(T_{\mu}))$ for some
$i=1,2$.  Since every node in $T_{1}$ and $T_{2}$ has in-degree at
most 1, every $[a]\in V(T_{po})$ has in-degree at most 2, and if it is
2, then $[a]$ must be of the first kind. 

Let $\ell_{i}:V(T_{i})\to V(T_{po})$, $i=1,2$, denote the inclusion
$V(T_{i})\hookrightarrow V(T_{1})\sqcup V(T_{2})$ followed by the
quotient mapping $V(T_{1})\sqcup V(T_{2})\to (V(T_{1})\sqcup
V(T_{2}))/\theta$: that is, $\ell_{i}(x)=[x]$ for every $x\in V(T_{i})$.
Note that, by construction,
$$
V(T_{po})=\ell_{1}(V(T_{1}))\cup \ell_{2}(V(T_{2}))
$$
and
$$
\ell_{1}(V(T_{1}))\cap \ell_{2}(V(T_{2}))=\ell_{1}(m_{1}(V(T_{\mu})))=
\ell_{2}(m_{2}(V(T_{\mu}))).
$$
It is straightforward to check that these mappings $\ell_{i}$ are
injective, satisfy that $\ell_{1}\circ m_{1}=\ell_{2}\circ m_{2}$, and
they define \emph{morphisms of graphs} $\ell_{i}:T_{i}\to T_{po}$,
$i=1,2$, in the sense that if $(a,b)\in E(T_{i})$, then
$(\ell_{i}(a),\ell_{i}(b))\in E(T_{po})$.

We shall call this graph $T_{po}$, together with these injective
morphisms $\ell_{i}:T_{i}\to T_{po}$, $i=1,2$, the \emph{join} of
$T_{1}$ and $T_{2}$ obtained through $m_{1}$ and $m_{2}$.

\begin{lemma} \label{lem-paths-po1}
Let $T_{1}$ and $T_{2}$ be trees, let $T_{\mu}$ be a largest common
$\CC_{*}$-subtree of $T_{1}$ and $T_{2}$, let $m_{1}:T_{\mu}\to T_{1}$
and $m_{2}:T_{\mu}\to T_{2}$ be any $\CC_{*}$-embeddings, and let
$T_{po}$ be the join of $T_{1}$ and $T_{2}$ obtained through $m_{1}$
and $m_{2}$.
\begin{enumerate}
\renewcommand{\labelenumi}{(\roman{enumi})}

\item If $r$ is the root of $T_{\mu}$, then $m_{1}(r)$ is the root of
$T_{1}$ or $r_{2}$ is the root of $T_{2}$.

\item For every $x,y\in V(T_{\mu})$, if $T_{po}$ contains a path from
$[m_{1}(x)]=[m_{2}(x)]$ to $[m_{1}(y)]=[m_{2}(y)]$, then $T_{\mu}$
contains a path from $x$ to $y$.

\item $T_{po}$ contains no circuit.
\end{enumerate}
\end{lemma}

\begin{proof}
(i) Assume that both $m_{1}(r)$ and $m_{2}(r)$ have parents, say
$v_{1}$ and $v_{2}$, respectively.  Lemma~\ref{lem-path} implies that
$v_{i}\notin m_{i}(V(T_{\mu}))$, for each $i=1,2$: otherwise, there
would be an arc in $T_{\mu}$ from the preimage of $v_{i}$ to $r$.
Then, we can enlarge $T_{\mu}$ by adding a new
node $r_{0}$ and an arc $(r_{0},r)$ and we can extend $m_{1}$ and
$m_{2}$ to this new tree by sending $r_{0}$ to $v_{1}$ and $v_{2}$,
respectively. In this way we obtain a tree strictly larger than
$T_{\mu}$ and $\CC_{*}$-embeddings of this new tree into $T_{1}$ and  
$T_{2}$,
against the assumption that $T_{\mu}$ is a largest $\CC_{*}$-subtree
of them.

(ii) We shall prove that if $T_{po}$ contains a path $[m_{1}(x)]\cami
[m_{1}(y)]$, then $T_{\mu}$ contains a path $x\cami y$, by induction
on its number $n$ of intermediate nodes in
$\ell_{1}(m_{1}(V(T_{\mu})))=\ell_{2}(m_{2}(V(T_{\mu})))$.

If $n=0$, that is, if no intermediate node in the path $[m_{1}(x)]\cami
[m_{1}(y)]$ comes from a
node of $T_{\mu}$, then all intermediate nodes come only from one of
the trees $T_{1}$ or $T_{2}$: assume, to fix ideas, that they come
from $T_{1}$, and that this path is
$$
([m_{1}(x)],[v_{1}],\ldots,[v_{k}],[m_{1}(y)]),
$$
with $[v_{1}],\ldots,[v_{k}]\in
\ell_{1}(V(T_{1}))\allowbreak -\ell_{1}(m_{1}(V(T_{\mu})))$.  Since
the nodes
belonging to $\ell_{1}(V(T_{1}))-\ell_{1}(m_{1}(V(T_{\mu})))$ are (as
equivalence classes) singletons, and an arc in $T_{po}$ involving one
node of this set must be induced by an arc in $E(T_{1})$, we conclude
that there exists a path
$$
(m_{1}(x),v_{1},\ldots,v_{k},m_{1}(y))
$$
in $T_{1}$.  Since $m_{1}$ is a $\CC_{*}$-embedding, and in particular
a minor embedding, by Lemma \ref{lem-path} this implies that there
exists a path  $x\cami y$ in $T_{\mu}$.

As the induction hypothesis, assume that the claim is true for paths
in $T_{po}$ with $n$ intermediate nodes in
$\ell_{1}(m_{1}(V(T_{\mu})))=\ell_{2}(m_{2}(V(T_{\mu})))$, and assume
now that the path  $[m_{1}(x)]\cami [m_{1}(y)]$ has $n+1$ such nodes.
Let $[m_{1}(a)]$
be the first intermediate node of this path belonging to
$\ell_{1}(m_{1}(V(T_{\mu})))$.  Then, by the case $n=0$, there is a
path $x\cami a$ in
$T_{\mu}$, and by the induction hypothesis there is
a path  $a\cami y$; by concatenating them we obtain the
path  $x\cami y$ in $T_{\mu}$ we were looking for.

(iii) Assume that $T_{po}$ contains a circuit.  If at most one node in
this circuit belongs to $\ell_{1}(m_{1}(V(T_{\mu})))$, then, arguing
as in the proof of (ii), we conclude that all arcs in this circuit are
induced by arcs in the same tree $T_{1}$ or $T_{2}$, and they would
form a circuit in this tree, which is impossible.  Therefore, two
different nodes in this circuit must belong to
$\ell_{1}(m_{1}(V(T_{\mu})))$.  This implies that there exist $x,y\in
V(T_{\mu})$, $x\neq y$, such that $T_{po}$ contains a path
$[m_{1}(x)]\cami [m_{1}(y)]$ and a path $[m_{1}(y)]\cami [m_{1}(x)]$.
By point (ii), this implies that $T_{\mu}$ contains a path $x\cami y$
and a path $y\cami x$, and hence a circuit, which is impossible.
Therefore, $T_{po}$ cannot contain any circuit.  \qed
\end{proof}

\begin{proposition}\label{lem-paths-po2}
Let $T_{1}$ and $T_{2}$ be trees, let $T_{\mu}$ be a largest common
$\CC_{*}$-subtree of $T_{1}$ and $T_{2}$, let $m_{1}:T_{\mu}\to T_{1}$
and $m_{2}:T_{\mu}\to T_{2}$ be any $\CC_{*}$-embeddings, and let
$T_{po}$ be the join of $T_{1}$ and $T_{2}$ obtained through $m_{1}$
and $m_{2}$.

\begin{enumerate}
\renewcommand{\labelenumi}{(\roman{enumi})}

\item For every $v,w\in V(T_{po})$, if $(v,w)\in E(T_{po})$ and there
is another path $v\cami w$ in $T_{po}$, then $v,w\in
\ell_{1}(V(T_{1}))\cap \ell_{2}(V(T_{2}))$, this path is unique, it is
a $\CC_{*}$-path and it has no intermediate node in
$\ell_{1}(V(T_{1}))\cap \ell_{2}(V(T_{2}))$.  In particular, if
$\CC_{*}$ is $\CC_{iso}$, then this situation cannot happen.

\item For every $v,w\in V(T_{po})$, if there are two different paths
from $v$ to $w$ in $T_{po}$ without any common intermediate node, then
one of them is the arc $(v,w)$, and then (ii) applies.  In particular,
again, this situation cannot happen if $\CC_{*}$ is $\CC_{iso}$.
\end{enumerate}
\end{proposition}

\begin{proof}
(i) If $(v,w)\in E(T_{po})$, then there exist, say, $a,b\in V(T_{1})$
such that $v=[a]$, $w=[b]$, and $(a,b)\in E(T_{1})$.  Assume now that
there is another path from $[a]$ to $[b]$ in $T_{po}$.  Since $T_{po}$
does not contain circuits by point (iii) in the last lemma, this path
cannot contain $[a]$ or $[b]$ as an intermediate node, and therefore
its first intermediate node is different from $[b]$ and its last
intermediate node is different from $[a]$.  In particular, $[b]$ has
in-degree 2 in $T_{po}$.

This implies that there exists some $y\in V(T_{\mu})$ such that
$b=m_{1}(y)$ and there exists some $c\in V(T_{2})$ such that
$(c,m_{2}(y))\in E(T_{2})$, and that there is a non-trivial path in
$T_{po}$ from $[a]$ to $[c]$.  Since $a\in V(T_{1})$ and $c\in
V(T_{2})$, this path must contain some node belonging to
$\ell_{1}(V(T_{1}))\cap \ell_{2}(V(T_{2}))$.  If it is not $[a]$, then
let $[m_{1}(x)]$ be the first intermediate node in the path $[a]\cami
[c]$ coming from $T_{\mu}$.  Since, in this case, $a\in
V(T_{1})-m_{1}(V(T_{\mu}))$, all intermediate nodes in the path
$[a]\cami [m_{1}(x)]$ come also from $V(T_{1})-m_{1}(V(T_{\mu}))$, and
therefore there exists a path $a\cami m_{1}(x)$ in $T_{1}$.  But, on
the other hand, since there is a path $[m_{1}(x)]\cami [m_{1}(y)]$ in
$T_{po}$ (consisting of the path $[m_{1}(x)]\cami [c]$ followed by the
arc $([c],[m_{1}(y)])$), from Lemma \ref{lem-paths-po1}.(ii) we deduce
that there exists a path $x\cami y$ in $T_{\mu}$ and hence a path
$m_{1}(x)\cami m_{1}(y)=b$ in $T_{1}$.  Summarizing, if $a\notin
m_{1}(V(T_{\mu}))$, then $T_{1}$ contains both an arc from $a$ to
$m_{1}(y)$ and a non-trivial path from $a$ to $m_{1}(y)$ (through
$m_{1}(x)$), which is impossible.

So, $a=m_{1}(x)$ for some $x\in V(T_{\mu})$.  Since
$(m_{1}(x),m_{1}(y))\in E(T_{1})$, Lemma \ref{lem-path} implies that
$(x,y)\in E(T_{\mu})$, and therefore there exists a $\CC_{*}$-path in
$T_{2}$ from $m_{2}(x)$ to $m_{2}(y)$ without any intermediate node in
$m_{2}(V(T_{\mu}))$: the uniqueness of paths in trees implies that this
path contains $c$ as its last intermediate node before $m_{2}(y)$.  To
begin with, this already shows that the situation considered in this
point cannot happen if $\CC_{*}$ is $\CC_{iso}$: a $\CC_{iso}$-path is
an arc, and therefore it does not contain any intermediate node.

Thus, we assume now that $\CC_{*}$ is $\CC_{min}$, $\CC_{hom}$ or
$\CC_{top}$.  The $\CC_{*}$-path $m_{2}(x)\cami m_{2}(y)$ in $T_{2}$
without any intermediate node in $m_{2}(V(T_{\mu}))$ and containing
$c$ as the last intermediate node induces a path from $[m_{2}(x)]=[a]$
to $[b]$ in $T_{po}$ containing $[c]$ and with all its intermediate
nodes in $\ell_{2}(V(T_{2}))-\ell_{2}(m_{2}(V(T_{\mu})))$.  This path
is a $\CC_{*}$-path.  If $\CC_{*}$ stands for $\CC_{min}$ or
$\CC_{top}$, it is obvious, because in these cases $\CC_{*}$-paths are
simply paths.  If $\CC_{*}$ is $\CC_{hom}$, then all intermediate
nodes in the path $m_{2}(x)\cami m_{2}(y)$ in $T_{2}$ have only one
child, and since they belong to $V(T_{2})-m_{2}(V(T_{\mu}))$ and
therefore they are not identified with any node from $T_{1}$, their
equivalence classes in $T_{po}$ have also out-degree 1, and hence the
path $[m_{2}(x)]\cami [m_{2}(y)]$ in $T_{po}$ it induces is also
elementary.

This proves that $v,w\in \ell_{1}(V(T_{1}))\cap \ell_{2}(V(T_{2}))$
and that, besides the arc $(v,w)$, there exists a  $\CC_{*}$-path  
$v\cami w$
without any intermediate node in $\ell_{1}(V(T_{1}))\cap
\ell_{2}(V(T_{2}))$, which contains $[c]$.  Assume finally that there
exists a ``third'' path from $v$ to $w$ other than the arc and this
$\CC_{*}$-path.  Since $w$ has in-degree at most 2 and $T_{po}$
contains no circuit, arguing as in the first paragraph of this proof
we deduce that this path consists of a path from $v$ to $[c]$ followed
by the arc $([c],w)$.  But $[c]$ has in-degree 1 in $T_{po}$, as well
as all intermediate nodes in the path from $v$ to $[c]$ induced by the
path $m_{2}(x)\cami c$ in $T_{2}$.  Therefore, this is the only path in
$T_{po}$ from $v$ to $[c]$.  This shows that there is only one path
from $v$ to $w$ in $T_{po}$ other than the arc $(v,w)$, and it is the
$\CC_{*}$-path without any intermediate node in
$\ell_{1}(V(T_{1}))\cap \ell_{2}(V(T_{2}))$ obtained above.

(ii) Assume that there exist two different paths from $v$ to $w$
without any intermediate node in common, and let $v_{1}$ and $v_{2}$ be
the nodes that precede $w$ in each one of these two paths; by
assumption $v_{1}\neq v_{2}$ and $(v_{1},w),(v_{2},w)\in E(T_{po})$.
Then, $w$ has in-degree 2 in $T_{po}$, and this implies that there
exist $y\in V(T_{\mu})$, $b\in V(T_{1})$ and $c\in V(T_{2})$ such
that, say, $v_{1}=[b]$, $v_{2}=[c]$, $w=[m_{1}(y)]=[m_{2}(y)]$, and
$(b,m_{1}(y))\in E(T_{1})$, $(c,m_{2}(y))\in E(T_{2})$.  By Lemma
\ref{lem-paths-po2}.(i), $y$ has a parent $x$ in $T_{\mu}$, and then
there are $\CC_{*}$-paths $m_{1}(x)\cami m_{1}(y)$ in $T_{1}$ and
$m_{2}(x)\cami m_{2}(y)$ in $T_{2}$.  This yields, up
to symmetry, three possibilities:

\begin{itemize}
\item If $m_{1}(x)=b$ and $m_{2}(x)=c$, then $[b]=[c]$, against the
assumption $v_{1}\neq v_{2}$.  Besides, if $\CC_{*}=\CC_{iso}$, then,
since $\CC_{iso}$-paths are arcs, it must happen that $m_{1}(x)=b$ and
$m_{2}(x)=c$.  So, if $\CC_{*}=\CC_{iso}$, the situation described in
the point we are proving cannot happen.  In the remaining two cases we
understand that $\CC_{*}\neq\CC_{iso}$.

\item If $m_{1}(x)=b$ and $m_{2}(x)\neq c$, then in $T_{po}$ we have
on the one hand the arc $([b],w)$ and on the other hand a path
$[b]\cami w$ induced by the path $m_{2}(x)\cami m_{2}(y)$ in $T_{2}$:
since $c$ is the parent of $m_{2}(y)$ in $T_{2}$, it is the last
intermediate node in the path $m_{2}(x)\cami m_{2}(y)$, and therefore
$[c]$ is the last intermediate node in the path $[b]\cami w$ induced by
$m_{2}(x)\cami m_{2}(y)$.  By (i), these are the only two paths from
$[b]$ to $w$.

Let us prove now that the path $v\cami w$ containing $[c]$ also
contains $[b]$.  Assume first that this path contains some node in
$\ell_{2}(m_{2}(T_{\mu}))$ other than $w$, and let $[m_{2}(z)]$ be the
last such node before $w$.  This means that $T_{po}$ contains a path
$[m_{2}(z)]\cami [m_{2}(y)]$ and therefore, by Lemma
\ref{lem-paths-po2}.(ii), there is a path $z\cami y$ in $T_{\mu}$.
But then this path must contain the parent $x$ of $y$, which implies
that the path $[m_{2}(z)]\cami [m_{2}(y)]$, and hence the path $v\cami
w$ through $[c]$, contains in this case $[b]=[m_{2}(x)]$.  

Assume now that the path $v\cami w$ containing $[c]$ does not contain
any node in $\ell_{2}(m_{2}(T_{\mu}))$ other than $w$.  Since $c\in
V(T_{2})$, this would mean that this path is completely induced by a
path in $T_{2}$, that is, $v=[a]$ for some $a\in
V(T_{2})-m_{2}(V(T_{\mu}))$ and there exists a path
$(a,\ldots,c,m_{2}(y))$ in $T_{2}$ with no intermediate node in
$m_{2}(V(T_{\mu}))$.  In this case, since there is a path
$m_{2}(x)\cami m_{2}(y)$ in $T_{2}$ and $m_{2}(x)$ is not contained in
the path $a\cami m_{2}(y)$, there would exist a non-trivial path
$m_{2}(x)\cami a$, which would induce a path from $[b]=[m_{2}(x)]$ to
$v=[a]$ forming a circuit with the path $v\cami [b]$. So, this case 
cannot happen.

% The equivalence classes of all these
% nodes have in-degree 1 in $T_{po}$, and then this path induces a path
% $v\cami [c]$ in $T_{po}$ with all nodes of in-degree 1.  Since there
% is a path $[b]\cami [c]$ in this graph and there is no non-trivial
% path $[b]\cami v$ (because there is a path $v\cami [b]$ and $T_{po}$
% does not contain circuits), it must happen that the path $v\cami [c]$
% in $T_{po}$ induced by the path $a\cami c$ contains $[b]$.

So, the path $v\cami w$ containing $[c]$ also contains $[b]$.  But, by
assumption, the paths $v\cami w$ containing $[b]$ and $[c]$ have no
common intermediate node.  Therefore, it must happen that $v=[b]$, and
hence one of the paths from $v$ to $w$ is an arc, as it is claimed in
the statement.

\item If $m_{1}(x)\neq b$ and $m_{2}(x)\neq c$, then there are
$\CC_{*}$-paths $(m_{1}(x),\ldots,b, m_{1}(y))$ and
$(m_{2}(x),\ldots,c, m_{2}(y))$ in $T_{1}$ and $T_{2}$, respectively,
without intermediate nodes coming from $V(T_{\mu})$.

In this case, we can enlarge $T_{\mu}$ by adding a new node $x_{0}$ and
replacing the arc $(x,y)$ by two arcs $(x,x_{0})$ and $(x_{0},y)$, and
we can extend $m_{1}$ and $m_{2}$ to this new node by sending it,
respectively, to $b$ and $c$.  It is clear that this new tree is
strictly larger than $T_{\mu}$.  Moreover, the extensions of $m_{1}$
and $m_{2}$ are $\CC_{*}$-embeddings: the new arc $(x,x_{0})$ is
transformed under them into the $\CC_{*}$-paths ---without
intermediate nodes coming from $V(T_{\mu})$--- that go from $m_{1}(x)$
to $b$ and from $m_{2}(x)$ to $c$, respectively; the new arc
$(x_{0},y)$ is transformed under them into the arcs $(b,m_{1}(y))$ and
$(c,m_{2}(y))$, respectively; and it is clear that if $m_{1}$ and
$m_{2}$ were topological embeddings, then their extensions are still
so, because the new node $x_{0}$ has only one child.  Thus, in this
way we obtain a new common $\CC_{*}$-subtree of $T_{1}$ of $T_{2}$
that is strictly larger than $T_{\mu}$, which yields a contradiction.
\end{itemize}
Summarizing, if $\CC_{*}$ is $\CC_{iso}$, then there cannot exist two
different paths $v\cami w$, and if $\CC_{*}$ is $\CC_{hom}$,
$\CC_{top}$, or $\CC_{iso}$, there can exist two different paths
$v\cami w$ without common intermediate nodes, but then the only case
that does not yield a contradiction is when one of these paths is an
arc.  \qed
\end{proof}

Let now $T_{\sigma}$ be the graph obtained from $T_{po}$ by removing
every arc that is subsumed by a path: that is, we remove from $T_{po}$
each arc $(v,w)$ for which there is another path $v\cami w$ in
$T_{po}$.  Note in particular that $V(T_{\sigma})=V(T_{po})$.  We
shall call this graph the \emph{$\CC_{*}$-sum} of $T_{1}$ and $T_{2}$
obtained through $m_{1}$ and $m_{2}$.

As a direct consequence of Lemma~\ref{lem-paths-po2}.(i), we have that
if $\CC_{*}=\CC_{iso}$, then $T_{\sigma}=T_{po}$, because if $(v,w)\in
E(T_{po})$, there does not exist any other path $v\cami w$, and
therefore no arc is removed from $T_{po}$ in the construction of
$T_{\sigma}$.  In the other three categories, still by
Lemma~\ref{lem-paths-po2}.(i) and its proof, if the arc $([a],[b])$
induced by an arc, say, $(a,b)\in E(T_{1})$ is removed because of the
existence of a second path $[a]\cami [b]$, then $a,b\in
m_{1}(T_{\mu})$, this second path is a $\CC_{*}$-path, and all its
intermediate nodes are equivalence classes of nodes in
$V(T_{2})-m_{2}(V(T_{\mu}))$.  In particular, since the arcs $(v,w)$
removed in the construction of $T_{\sigma}$ are such that $v,w\in
\ell_{1}(m_{1}(V(T_{\mu})))=\ell_{2}(m_{2}(V(T_{\mu})))$ and the
$\CC_{*}$-paths that make these arcs to be removed have no
intermediate node in this set, these paths are not modified in the
construction of $T_{\sigma}$, and the arcs can be removed in any
order.

%%%

\begin{proposition} \label{th-supertree-gral}
For every two trees $T_{1}$ and $T_{2}$, any $\CC_{*}$-sum of $T_{1}$
and $T_{2}$ is a common $\CC_{*}$-su\-per\-tree of them.
\end{proposition}

\begin{proof}
Let $T_{1}$ and $T_{2}$ be two trees, let $T_{\mu}$ be a largest
common $\CC_{*}$-subtree of them and let $m_{1}:T_{\mu}\to T_{1}$ and
$m_{2}:T_{\mu}\to T_{2}$ be any $\CC_{*}$-embeddings. Let $T_{\sigma}$
be the $\CC_{*}$-sum of $T_{1}$ and $T_{2}$ obtained through $m_{1}$
and $m_{2}$, and let $\ell_{i}:V(T_{i})\to
V(T_{\sigma})=(V(T_{1})\sqcup V(T_{2}))/\theta$, $i=1,2$, stand for
the corresponding restrictions of the quotient mappings.

Every arc removed from the join $T_{po}$ of $T_{1}$ and $T_{2}$ in the
construction of $T_{\sigma}$ is subsumed by a path in $T_{po}$.  This
implies that, for every $x,y\in V(T_{po})$, there is a path $x\cami y$
in $T_{po}$ if and only if there is a path $x\cami y$ in $T_{\sigma}$.
In particular, since the only nodes in $T_{po}$ than can possibly have
no parent are the images of the roots of $T_{1}$ and $T_{2}$, the same
also happens in $T_{\sigma}$.

Now, by Lemma \ref{lem-paths-po1}.(i), if $r$ is the root of
$T_{\mu}$ then $m_{1}(r)$ is the root $r_{1}$ of $T_{1}$ or
$m_{2}(r)$ is the root $r_{2}$ of $T_{2}$.  If
$m_{1}(r)=r_{1}$ and $m_{2}(r)=r_{2}$, then
$[r_{1}]=[r_{2}]$ is the only node in $T_{\sigma}$ without parent, and
every node $v$ in $T_{po}$ (as well as in $T_{\sigma}$, as we said)
can be reached from this node through a path: if $v=[a_{1}]$, with
$a_{1}\in V(T_{1})$, through the image of the path $r_{1}\cami a_{1}$
in $T_{1}$, and if $v=[a_{2}]$, with $a_{2}\in V(T_{2})$, through the
image of the path $r_{2}\cami a_{2}$ in $T_{2}$.  If, on the contrary,
say, $m_{1}(r)=r_{1}$ but $m_{2}(r)\neq r_{2}$, then
$[r_{2}]$ is the only node in $T_{\sigma}$ with no parent and every
node in $T_{\sigma}$ can be reached from this node through a path:
every node of the form $[a_{2}]$, with $a_{2}\in V(T_{2})$, through
the image of the path $r_{2}\cami a_{2}$ in $T_{2}$, and every node of
the form $[a_{1}]$, with $a_{1}\in V(T_{1})$, through the path
obtained by concatenating the image of the path $r_{2}\cami
m_{2}(r)$ in $T_{2}$ and the image of the path $r_{1}\cami
a_{1}$ in $T_{1}$.

Thus, $T_{\sigma}$ has one, and only one, node without parent, and
every other node in $T_{\sigma}$ can be reached from it through a
path.  Moreover, every node in $T_{\sigma}$ has in-degree at most 1.
Indeed, if a node $w$ has in-degree 2 in $T_{po}$, say
$(v_{1},w),(v_{2},w)\in E(T_{po})$, then there will exist some node
$v$ and paths $v\cami v_{1}$ and $v\cami v_{2}$ with no common
intermediate node.  But then, by Lemma \ref{lem-paths-po2}.(ii), $v$
will be one of the nodes $v_{1}$ or $v_{2}$, say $v=v_{1}$, and then
the arc $(v_{1},w)\in E(T_{po})$ is subsumed by the path $v_{1}\cami
w$ through $v_{2}$, and hence it is removed in the construction of
$T_{\sigma}$, leaving only the arc $(v_{2},w)$.  So, every node in
$T_{\sigma}$ has in-degree at most 1, and it can be reached through a
path from the only node without a parent. This proves that $T_{\sigma}$ is a tree.

Now we have to prove that $\ell_{1}:T_{1}\to T_{\sigma}$ and
$\ell_{2}:T_{2}\to T_{\sigma}$ are $\CC_{*}$-embeddings.  We shall
prove that $\ell_{1}$ is a $\CC_{*}$-embedding.  Recall that the
mapping $\ell_{1}: V(T_{1})\to V(T_{po})=V(T_{\sigma})$ is injective,
and note that, by Lemma \ref{lem-paths-po2}, if $(a,b)\in E(T_{1})$,
then there is a $\CC_{*}$-path in $T_{\sigma}$ from $\ell_{1}(a)=[a]$
to $\ell_{1}(b)=[b]$ that does not contain any intermediate node in
$\ell_{1}(V(T_{1}))\cap \ell_{2}(V(T_{2}))$: either the arc 
$([a],[b])$ induced by
the arc in $T_{1}$ or the $\CC_{*}$-path $[a]\cami [b]$ that made this arc to be
removed.  This shows that $\ell_{1}$ is a $\CC_{*}$-embedding when
$\CC_{*}$ is $\CC_{iso}$, $\CC_{hom}$ or $\CC_{min}$.

In the case of $\CC_{top}$, it remains to prove that if
$(a,b),(a,c)\in E(T_{1})$, then the paths $[a]\cami [b]$ and $[a]\cami
[c]$ are divergent. Up to symmetry, there are three possibilities to
discuss:
\begin{itemize}
\item If the paths $[a]\cami [b]$ and $[a]\cami [c]$ are both arcs,
then the injectivity of $\ell_{1}$ implies that they are different and
therefore they define divergent paths.

\item If the path $[a]\cami [b]$ is an arc and the path $[a]\cami [c]$
has intermediate nodes, and if they did not diverge, $[b]$ would be
the first intermediate node of the path $[a]\cami [c]$.  But this is
impossible, because, since the arc $([a],[c])\in E(T_{po})$ has been
removed in the construction of $T_{\sigma}$, all intermediate nodes of
the path $[a]\cami [c]$ are equivalence classes of nodes in
$V(T_{2})-m_{2}(V(T_{\mu}))$.

\item If both paths $[a]\cami [b]$ and $[a]\cami [c]$ have
intermediate nodes, then both arcs $([a],[b]),([a],\allowbreak [c])\in E(T_{po})$
have been removed in the construction of $T_{\sigma}$, and therefore
there are $x,y,z\in V(T_{\mu})$ such that $(x,y),(x,z)\in E(T_{\mu})$,
$m_{1}(x)=a$, $m_{1}(y)=b$, $m_{1}(z)=c$, and the intermediate nodes
of the paths $[a]\cami [b]$ and $[a]\cami [c]$ are the equivalence
classes of the intermediate nodes of the paths $m_{2}(x)\cami
m_{2}(y)$ and $m_{2}(x)\cami m_{2}(z)$ in $T_{2}$.  Now, since $m_{2}$
is a topological embedding, these paths $m_{2}(x)\cami m_{2}(y)$ and
$m_{2}(x)\cami m_{2}(z)$ have no common intermediate node.  Since
$\ell_{2}$ is injective, no image of an intermediate node of the path
$m_{2}(x)\cami m_{2}(y)$ is equal to the image of an intermediate node
of the path $m_{2}(x)\cami m_{2}(z)$, and thus the paths $[a]\cami
[b]$ and $[a]\cami [c]$ are divergent.
\end{itemize}
Therefore, if $\CC_{*}=\CC_{top}$, $\ell_{1}$  is a topological
embedding.\qed
\end{proof}

Theorem \ref{th-po} below extends the last proposition in the
algebraic direction, by showing that $\CC_{*}$-sums are not only
common $\CC_{*}$-subtrees, but pushouts. In its proof we shall use
several times the following technical fact, which we establish first as  
a lemma.

\begin{lemma}\label{lem-previ-th-po}
Let $\CC_{*}$ be $\CC_{hom}$, $\CC_{top}$ or $\CC_{min}$.  Let $T_{1}$
and $T_{2}$ be trees, let $T_{\mu}$ be a largest common
$\CC_{*}$-subtree of $T_{1}$ and $T_{2}$, let $m_{1}:T_{\mu}\to T_{1}$
and $m_{2}:T_{\mu}\to T_{2}$ be any $\CC_{*}$-embeddings, and let
$f_{1}:T_{1}\to T$ and $f_{2}:T_{2}\to T$ be any $\CC_{*}$-embeddings
such that $f_{1}\circ m_{1}=f_{2}\circ m_{2}$.

There do not exist $x\in V(T_{\mu})$, $p\in
V(T_{1})-m_{1}(V(T_{\mu}))$, and $q\in V(T_{2})-m_{2}(V(T_{\mu}))$
such that $(m_{1}(x),p)\in E(T_{1})$, $(m_{2}(x),q)\in E(T_{2})$,
and $f_{1}(p)$ and $f_{2}(q)$ are connected by a path.
\end{lemma}

\begin{proof}
Assume that there exist $x\in V(T_{\mu})$, $p\in
V(T_{1})-m_{1}(V(T_{\mu}))$, and $q\in V(T_{2})-m_{2}(V(T_{\mu}))$
such that $(m_{1}(x),p)\in E(T_{1})$, $(m_{2}(x),q)\in E(T_{2})$, and
there is, say, a path $f_{2}(q)\cami f_{1}(p)$, in such a way that
$f_{2}(q)$ is an intermediate node in the path $f_{1}(m_{1}(x))\cami
f_{1}(p)$. We shall look for a contradiction.

Under these assumptions, we can enlarge $T_{\mu}$ by adding a new node
$y$, a new arc $(x,y)$, and replacing by a new arc $(y,z)$ every arc
$(x,z)$ such that the path $m_{1}(x)\cami m_{1}(z)$ in $T_{1}$
contains $p$.  It is clear that the graph $\hat{T}_{\mu}$ obtained in
this way is a tree, strictly larger than $T_{\mu}$.  We can extend
$m_{1}$ and $m_{2}$ to $\hat{T}_{\mu}$ by defining $m_{1}(y)=p$ and
$m_{2}(y)=q$.  If we prove that the mappings
$m_{1}:V(\hat{T}_{\mu})\to V(T_{1})$ and $m_{2}:V(\hat{T}_{\mu})\to
V(T_{2})$ defined in this way are $\CC_{*}$-embeddings
$m_{1}:\hat{T}_{\mu}\to T_{1}$ and $m_{2}:\hat{T}_{\mu}\to T_{2}$,
this will contradict the assumption that $T_{\mu}$ is a largest common
$\CC_{*}$-subtree of $T_{1}$ and $T_{2}$.

Now, on the one hand, the arc $(x,y)$
is transformed under $m_{1}$ and $m_{2}$ into the arcs $(m_{1}(x),p)$  
and
$(m_{1}(x),q)$, respectively.  Assume now that $\hat{T}_{\mu}$
contains a new arc $(y,z)$.  This means that $T_{\mu}$ contained
$(x,z)$ and that $p$ is the first intermediate node of the
$\CC_{*}$-path $m_{1}(x)\cami m_{1}(z)$, which does not have any
intermediate node in $m_{1}(V(T_{\mu}))$.  This implies that there
exists in $T_{1}$ a $\CC_{*}$-path without intermediate nodes in
$m_{1}(V(\hat{T}_{\mu}))$ from $p=m_{1}(y)$ to $m_{1}(z)$.  As far
as $m_{2}$ goes, note that the arc $(x,z)$ in $T_{\mu}$ induces
under $f_{1}\circ m_{1}$ a $\CC_{*}$-path from
$f_{1}(m_{1}(x))=f_{2}(m_{2}(x))$ to
$f_{1}(m_{1}(z))=f_{2}(m_{2}(z))$ that contains $f_{1}(p)$.  This path
also contains $f_{2}(q)$, because this node is contained in the path
from $f_{1}(m_{1}(x))=f_{2}(m_{2}(x))$ to $f_{1}(p)$. So, there 
exists a $\CC_{*}$-path $f_{2}(q)\cami f_{2}(m_{2}(z))$, which 
entails the existence of a $\CC_{*}$-path $q\cami m_{2}(z)$ in $T_{2}$.
And since this path is actually a piece of the path $m_{2}(x)\cami 
m_{2}(z)$, it has no intermediate node in $m_{2}(V(T_{\mu}))$.

This shows that $m_{1}:\hat{T}_{\mu}\to T_{1}$ and
$m_{2}:\hat{T}_{\mu}\to T_{2}$ transform arcs into $\CC_{*}$-paths
without any intermediate node coming from $\hat{T}_{\mu}$, and hence
that they are $\CC_{*}$-morphisms when $\CC_{*}$ is $\CC_{hom}$ or
$\CC_{min}$.  When $\CC_{*}=\CC_{top}$, it remains to check that
$m_{1}$ and $m_{2}$ transform pairs of arcs with the same source node
into divergent paths.  To do it, note first that in this case $x$ has 
at most one child $z$ such that the path
$m_{1}(x)\cami m_{1}(z)$ in $T_{1}$ contains $p$, because the
paths in $T_{1}$ from $m_{1}(x)$ to the images under $m_{1}$ of
the children of $x$ diverge. Therefore, the new node $y$ has
out-degree at most 1 in $\hat{T}_{\mu}$.  So, to prove that $m_{1}$
and $m_{2}$ are topological embeddings, it is enough to check
that if $y_{1}$ is any child of $x$ in $\hat{T}_{\mu}$ other than
$y$, the paths $m_{i}(x)\cami m_{i}(y)$ and
$m_{i}(x)\cami m_{i}(y_{1})$ in each $T_{i}$ diverge.  For
$i=1$ it is obvious, because the path $m_{1}(x)\cami m_{1}(y)$
is simply the arc $(m_{1}(x),p)$ and, by assumption, $p$ is not
contained in the path $m_{1}(x)\cami m_{1}(y_{1})$.  As far as
the case $i=2$ goes, the path $m_{2}(x)\cami m_{2}(y)$ is simply the
arc $(m_{2}(x),q)$, and thus it is enough to check that $q$ is not
contained in the path $m_{2}(x)\cami m_{2}(y_{1})$.  But the paths
from $f_{1}(m_{1}(x))=f_{2}(m_{2}(x))$ to $f_{1}(p)$ and to
$f_{1}(m_{1}(y_{1}))=f_{2}(m_{2}(y_{1}))$ diverge because $f_{1}$ is a
topological embedding, and therefore, since $f_{2}(q)$ is contained in
the fist one, it cannot be contained in the second one, which implies
that $q$ cannot be contained in the path $m_{2}(x)\cami
m_{2}(y_{1})$. This finishes the proof that, when $\CC_{*}=\CC_{top}$,
$m_{1}$ and $m_{2}$ are topological embeddings.\qed
\end{proof}

\begin{theorem} \label{th-po}
Let $T_{1}$ and $T_{2}$ be trees, let $T_{\mu}$ be a largest common
$\CC_{*}$-subtree of $T_{1}$ and $T_{2}$, and let $m_{1}:T_{\mu}\to
T_{1}$ and $m_{2}:T_{\mu}\to T_{2}$ be any $\CC_{*}$-embeddings.

Then, the $\CC_{*}$-sum $T_{\sigma}$ of $T_{1}$ and $T_{2}$ obtained
through $m_{1}$ and $m_{2}$, together with the $\CC_{*}$-embeddings
$\ell_{1}:T_{1}\to T_{\sigma}$ and $\ell_{2}:T_{2}\to T_{\sigma}$, is
a pushout in $\CC_{*}$ of $m_{1}$ and $m_{2}$.
\end{theorem}

%%%%%

\begin{proof}
It is clear that $\ell_{1}\circ m_{1}=\ell_{2}\circ m_{2}$.
Therefore, it remains to prove that $T_{\sigma}$, together with the
$\CC_{*}$-embeddings $\ell_{1}:T_{1}\to T_{\sigma}$ and
$\ell_{2}:T_{2}\to T_{\sigma}$, satisfies the universal property of
pushouts in $\CC_{*}$.

So, let $f_{1}:T_{1}\to T$ and $f_{2}:T_{2}\to T$ be any
$\CC_{*}$-embeddings such that $f_{1}\circ m_{1}=f_{2}\circ m_{2}$.
It is well-known that there exists one, and only one, mapping
$f:(V(T_{1})\sqcup V(T_{2}))/\theta\to V(T)$ such that $f\circ
\ell_{1}=f_{1}$ and $f\circ \ell_{2}=f_{2}$: namely, the one defined
by $f([a])=f_{1}(a)$ if $a\in V(T_{1})$ and $f([a])=f_{2}(a)$ if $a\in
V(T_{2})$.  We must prove that this mapping $f$ is a
$\CC_{*}$-embedding.

Let us prove first that it is injective.  Assume that there exist
$v,w\in V(T)$, $v\neq w$, such that $f(v)=f(w)$.  Since $f_{1}$ and
$f_{2}$ are injective, it is clear that they cannot be classes of
nodes of the same tree $T_{i}$.  Thus, there exist $a\in
V(T_{1})-m_{1}(V(T_{\mu}))$ and $b\in V(T_{2})-m_{2}(V(T_{\mu}))$ such
that $v=[a]$ and $w=[b]$ and $f_{1}(a)=f_{2}(b)$.

By Lemma \ref{lem-paths-po1}.(i), the image under some $m_{i}$ of the
root of $T_{\mu}$ is the root of the corresponding $T_{i}$.  This
implies that there exists a path from the image of a node in $T_{\mu}$
to one of these nodes $a$ or $b$ in the corresponding tree.  Moreover,
if there exists, say, some $x\in V(T_{\mu})$ such that there is a path
$m_{1}(x)\cami a$ in $T_{1}$, then there is a path from
$f_{1}(m_{1}(x))=f_{2}(m_{2}(x))$ to $f_{1}(a)=f_{2}(b)$ in $T$, and
hence a path $m_{2}(x)\cami b$ in $T_{2}$.  By symmetry, if there
exists some $x\in V(T_{\mu})$ such that there is a path $m_{2}(x)\cami
b$ in $T_{2}$, then there is a path $m_{1}(x)\cami a$ in $T_{1}$.

This shows that there exists a node $x_{0}\in V(T_{\mu})$ such that
there exist paths $m_{1}(x_{0})\cami a$ in $T_{1}$ and
$m_{2}(x_{0})\cami b$ in $T_{2}$ without any intermediate node in
$m_{1}(V(T_{\mu}))$ or $m_{2}(V(T_{\mu}))$, respectively.  These paths
induce, through $f_{1}$ and $f_{2}$, the same path from
$f_{1}(m_{1}(x_{0}))=f_{2}(m_{2}(x_{0}))$ to $f_{1}(a)=f_{2}(b)$ in
$T$ (because of the uniqueness of paths in trees).  Let now $e$ be the
child of $m_{1}(x_{0})$ contained in the path $m_{1}(x_{0})\cami a$ in
$T_{1}$, and $d$ the child of $m_{2}(x_{0})$ in the path
$m_{2}(x_{0})\cami b$ in $T_{2}$.  Then $f_{1}(e)$ and $f_{2}(d)$ are
contained in the path from $f_{1}(m_{1}(x_{0}))=f_{2}(m_{2}(x_{0}))$
to $f_{1}(a)=f_{2}(b)$ in $T$, and hence, they are connected by a
path.

When $\CC_{*}$ is $\CC_{hom}$, $\CC_{top}$ or $\CC_{min}$, Lemma
\ref{lem-previ-th-po} says that this situation is impossible, and
therefore $f$ must be injective.  In the case when
$\CC_{*}=\CC_{iso}$, since $f_{1}$ and $f_{2}$ transform arcs into
arcs, it must happen that $f_{1}(e)=f_{2}(d)$.  This allows us to
enlarge $T_{\mu}$, by adding a new node $y_{0}$ and a new arc
$(x_{0},y_{0})$: it is clear that the graph $\hat{T}_{\mu}$ obtained in
this way is a tree.  We extend $m_{1}$ and $m_{2}$ to $\hat{T}_{\mu}$
by defining $m_{1}(y_{0})=e$ and $m_{2}(y_{0})=d$.  The mappings
$m_{1}:V(\hat{T}_{\mu})\to V(T_{1})$ and $m_{2}:V(\hat{T}_{\mu})\to
V(T_{2})$ defined in this way are isomorphic embeddings
$m_{1}:\hat{T}_{\mu}\to T_{1}$ and $m_{2}:\hat{T}_{\mu}\to T_{2}$.
Indeed, they are injective because their restrictions to $T_{\mu}$ are
injective and, by assumption, $e\notin m_{1}(V(\hat{T}_{\mu}))$ and
$d\notin m_{2}(V(\hat{T}_{\mu}))$, and they transform arcs into arcs
because their restrictions to $T_{\mu}$ do so and
$(m_{i}(x_{0}),m_{i}(y_{0}))\in E(T_{i})$ for each $i=1,2$.  In this
way we obtain a common isomorphic subtree of $T_{1}$ and $T_{2}$ that
is strictly larger than $T_{\mu}$, which yields a contradiction.
Therefore, $f$ is also injective in this case.

So, $f:V(T_{\sigma})\to V(T)$ is always injective.  Now, assume
$(v,w)\in T_{\sigma}$.  Then, for some $i=1,2$, there exist $a,b\in
V(T_{i})$ such that $v=[a]$, $w=[b]$, and $(a,b)\in E(T_{i})$: to fix
ideas, assume that $i=1$.  This implies that there is a $\CC_{*}$-path
from $f(v)=f_{1}(a)$ to $f(w)=f_{1}(b)$ in $T$.  If
$\CC_{*}=\CC_{iso}$, this already proves that $f$ is an isomorphic
embedding.

Thus, henceforth, we shall assume that $\CC_{*}\neq \CC_{iso}$.  In this
case, we must check that no intermediate node of this $\CC_{*}$-path  
$f(v)\cami
f(w)$ belongs to $f(V(T_{\sigma}))=f_{1}(V(T_{1}))\cup
f_{2}(V(T_{2}))$.  Now, $f_{1}$ being a $\CC_{*}$-embedding, we
already know that no intermediate node of this path belongs to
$f_{1}(V(T_{1}))$, and therefore we only have to check that no
intermediate node belongs to $f_{2}(V(T_{2}))$, either.  Before
proceeding, note that we have already proved that $f$ sends arcs to
$\CC_{*}$-paths, and hence that this mapping transforms paths in
$T_{\sigma}$ into paths in $T$.

Assume that there is some $c\in V(T_{2})$ such that $f_{2}(c)$ is
an intermediate node of the path $f_{1}(a)\cami f_{1}(b)$ in $T$.
This prevents the existence of paths $[c]\cami [a]$ or $[b]\cami [c]$
in $T_{\sigma}$: the image of such a path under $f$ would be a path in
$T$ that would build up a circuit with the path from $f_{1}(a)=f([a])$
to $f_{2}(c)=f([c])$ or from $f_{2}(c)=f([c])$ to $f_{1}(b)=f([b])$,
respectively, that we already know to exist.  Moreover, $c\notin
m_{2}(V(T_{\mu}))$, because if $c\in m_{2}(V(T_{\mu}))$, then
$f_{2}(c)\in f_{2}(m_{2}(V(T_{\mu})))=f_{1}(m_{1}(V(T_{\mu})))
\subseteq f_{1}(V(T_{1}))$.

After excluding these possibilities, we still must discuss several
cases:

\begin{itemize}

\item $a=m_{1}(x)$ and $b=m_{1}(y)$ for some $x,y\in V(T_{\mu})$.  In
this case, by Lemma~\ref{lem-path}, the existence of an arc from
$\ell_{2}(m_{2}(x))=\ell_{1}(m_{1}(x))=[a]$ to
$\ell_{2}(m_{2}(y))=\ell_{1}(m_{1}(y))=[b]$ implies the existence of
an arc from $m_{2}(x)$ to $m_{2}(y)$ in $T_{2}$.  Since $f_{2}$ is a
$\CC_{*}$-embedding, the path from $f_{2}(m_{2}(x))=f_{1}(a)$ to
$f_{2}(m_{2}(y))=f_{1}(b)$ does not contain any intermediate
node in $f_{2}(V(T_{2}))$, which contradicts the
existence of $c$.

\item $a=m_{1}(x)$ for some $x\in V(T_{\mu})$, but $b\notin
m_{1}(V(T_{\mu}))$.  In this case, since $f_{2}$ is a
$\CC_{*}$-embedding, the existence of a $\CC_{*}$-path
$f_{2}(m_{2}(x))=f_{1}(a)\cami f_{2}(c)$ in $T$  implies, by Corollary
\ref{lem-path-2}, the existence of a $\CC_{*}$-path $m_{2}(x)\cami c$
in $T_{2}$.  And this path cannot have any intermediate node in
$m_{2}(V(T_{\mu}))$: any intermediate node in this set would become,
under $f_{2}$, an intermediate node in
$f_{2}(m_{2}(V(T_{\mu})))=f_{1}(m_{1}(V(T_{\mu}))) \subseteq
f_{1}(V(T_{1}))$ of the path $f_{2}(m_{2}(x))\cami f_{2}(c)$.  Let $d$
be the child of $m_{2}(x)$ contained in this path $m_{2}(x)\cami c$.
The path $f_{2}(m_{2}(x))=f_{1}(a)\cami f_{2}(c)$ contains
$f_{2}(d)$, and therefore $f_{2}(d)$ is an intermediate node of the
path $f_{1}(a)\cami f_{1}(b)$.  But then this situation is impossible
by Lemma \ref{lem-previ-th-po}.

\item $a\notin m_{1}(V(T_{\mu}))$.  Since, by Lemma
\ref{lem-paths-po1}.(i), the image under $m_{1}$ or  $m_{2}$ of
the root of $T_{\mu}$ is the root of $T_{1}$ or $T_{2}$, respectively,
we know that there exists some $x\in V(T_{\mu})$ such that there is a
path $m_{1}(x)\cami a$ in $T_{1}$ or a path $m_{2}(x)\cami c$ in
$T_{2}$.  It turns out that the existence of such a path
$m_{1}(x)\cami a$ in $T_{1}$ or $m_{2}(x)\cami c$ in $T_{2}$ implies
the existence of paths $m_{1}(x)\cami a$ \emph{and} $m_{2}(x)\cami c$
in $T_{1}$ and $T_{2}$, respectively.  Indeed, if there exists a path
$m_{1}(x)\cami a$, then there is a path $f_{1}(m_{1}(x))\cami
f_{1}(a)$ in $T$, which, composed with the path $f_{1}(a)\cami
f_{2}(c)$, yields a path $f_{2}(m_{2}(x))=f_{1}(m_{1}(x))\cami
f_{2}(c)$, and this path, on its turn, implies a path $m_{2}(x)\cami
c$ in $T_{2}$.  Conversely, if there exists a path $m_{2}(x)\cami c$,
then there is a path from $f_{2}(m_{2}(x))$ to $f_{2}(c)$ in $T$.
Since there is also a path $f_{1}(a)\cami f_{2}(c)$ and
$f_{2}(m_{2}(x))=f_{1}(m_{1}(x))$ cannot be an intermediate node of
the path $f_{1}(a)\cami f_{2}(c)$ (because this path does not contain
any intermediate node in $f_{1}(V(T_{1}))$), it must happen that
$f_{1}(a)$ is intermediate in the path $f_{2}(m_{2}(x))\cami
f_{2}(c)$, that is, that there is a path
$f_{1}(m_{1}(x))=f_{2}(m_{2}(x))\cami f_{1}(a)$ which, finally,
implies a path $m_{1}(x)\cami a$ in $T_{1}$.

So, we can take $x\in V(T_{\mu})$ such that, on the one hand, there
exist paths $m_{1}(x)\cami a$ {and} $m_{2}(x)\cami c$ in $T_{1}$ and
$T_{2}$ and, on the other hand, there do not exist paths $m_{1}(y)\cami
a$ in $T_{1}$ or $m_{2}(y)\cami c$ in $T_{2}$ for any child $y$ of it.
Let then $e$ be the child of $m_{1}(x)$ contained in the path
$m_{1}(x)\cami a$ in $T_{1}$, and  $d$  the child of $m_{2}(x)$
contained in the path $m_{2}(x)\cami c$ in $T_{2}$.  The uniqueness of
paths in $T$ implies that the path $f_{2}(m_{2}(x))\cami f_{2}(c)$,
which contains $f_{2}(d)$, is the concatenation of the path
$f_{1}(m_{1}(x))\cami f_{1}(a)$, which contains $f_{1}(e)$, and the
path $f_{1}(a)\cami f_{2}(c)$.  Therefore, $f_{1}(e)$ and $f_{2}(d)$
are connected by a path.  By Lemma \ref{lem-previ-th-po}, this
situation cannot happen.
\end{itemize}

Therefore, $f$ transforms arcs into $\CC_{*}$-paths without
intermediate nodes in $f(V(T_{\sigma}))$, and thus it is a
$\CC_{*}$-embedding when $\CC_{*}$ is $\CC_{hom}$ or $\CC_{min}$.
This proves the universal property of pushouts, and with it the
statement, for these categories.  It remains to prove it in
$\CC_{top}$.  

So far, we know that, if we are in $\CC_{top}$,
then $f$ transforms arcs into paths without intermediate nodes in
$f(V(T_{\sigma}))$.  Now we must prove that it transforms arcs with
the same source node into divergent paths.  So, assume there are arcs
$(v,w)$ and $(v,u)$ in $T_{\sigma}$ with $w\neq u$.

If these arcs are induced by arcs in the same tree, i.e., if there
exist $(a,b),\!(a,c)\!\in V(T_{i})$, for some $i=1,2$, such that  
$v=[a]$,
$w=[b]$ and $u=[c]$, then, since $f_{i}$ is a topological embedding,
the paths from $f(v)=f_{i}(a)$ to $f(w)=f_{i}(b)$ and to
$f(u)=f_{i}(c)$ are divergent.  Now consider the case when each one of
these arcs is induced by an arc in a different tree.  In this case,
there exist $x\in V(T_{\mu})$, $b\in V(T_{1})$ and $c\in V(T_{2})$
such that, say, $v=[m_{1}(x)]=[m_{2}(x)]$, $w=[b]$ and $u=[c]$, and  
there
are arcs $(m_{1}(x),b)\in E(T_{1})$ and $(m_{2}(x),c)\in E(T_{2})$.

If there exists $y\in V(T_{\mu})$ such that $m_{1}(y)=b$, then, by
Lemma \ref{lem-path}, $(x,y)\in E(T_{\mu})$ and hence there exists a
path $m_{1}(x)\cami m_{2}(y)$ in $T_{2}$.  But since there is an arc
from $[m_{2}(x)]=[m_{1}(x)]$ to $[m_{2}(y)]=[b]$ in $T_{\sigma}$, the
path $m_{2}(x)\cami m_{2}(y)$ in $T_{2}$ must also be an arc
(otherwise, it would induce a path in $T_{po}$ that would have made
the arc $(v,w)$ to be removed in the construction of $T_{\sigma}$).
Therefore, the arc $(v,w)$ is induced by the arc $(m_{2}(x),m_{2}(y))$
in $T_{2}$, and thus both arcs $(v,w)$ and $(v,u)$ are induced by arcs
in $T_{2}$ and the paths $f(v)\cami f(w)$ and $f(v)\cami f(u)$ are
divergent, as we have just seen.  In a similar way, if there exists
$y\in V(T_{\mu})$ such that $m_{2}(y)=c$, then both arcs $(v,w)$ and
$(v,u)$ are induced by arcs in $T_{1}$ and the paths $f(v)\cami f(w)$
and $f(v)\cami f(u)$ are divergent.

Consider finally the case when neither $b$ nor $c$ have a preimage in
$T_{\mu}$.  There are two possibilities to discuss:
\begin{itemize}
\item If there exists an arc $(x,z)\in V(T_{\mu})$ such that $b$ is
the first intermediate node of the path $m_{1}(x)\cami m_{1}(z)$, then
$w=[b]$ is  the first intermediate node of the path $[m_{1}(x)]\cami
[m_{1}(z)]$.  In particular, $u=[c]$ does not appear in this last path,
which implies that the arc $(m_{2}(x),c)$ and the path $m_{2}(x)\cami
m_{2}(z)$ are divergent.  Since $f_{2}$ is a topological embedding,
the paths in $T$ from $f_{2}(m_{2}(x))$ to $f_{2}(c)$ and from
$f_{2}(m_{2}(x))=f_{1}(m_{1}(x))$ to $f_{2}(m_{2}(z))=f_{1}(m_{1}(z))$
are also divergent.  Since $f_{1}(b)$ is contained in this
last path, we finally deduce that the paths from
$f(v)=f_{2}(m_{2}(x))=f_{1}(m_{1}(x))$ to $f(u)=f_{1}(b)$ and to
$f(w)=f_{2}(c)$ are divergent.

The case when there exists an arc $(x,z)\in V(T_{\mu})$ such that $c$
is the first intermediate node of the path  $m_{2}(x)\cami
m_{2}(z)$ is solved in a similar way.

\item If there is no arc $(x,z)$ in $T_{\mu}$ such that $b$ or $c$ are
intermediate nodes of the paths $m_{1}(x)\cami m_{1}(z)$ or
$m_{2}(x)\cami m_{2}(z)$, respectively, then we can enlarge $T_{\mu}$
by adding to it a new node $y_{0}$ and an arc $(x,y_{0})$, and we can
extend $m_{1}$ and $m_{2}$ to this new tree by defining
$m_{1}(y_{0})=b$ and $m_{2}(y_{0})=c$, and it is straightforward to
prove that in this way we obtain a topological subtree of $T_{1}$ and
$T_{2}$ strictly larger than $T_{\mu}$, which contradicts the
assumption that $T_{\mu}$ is a largest common topological subtree of
$T_{1}$ and $T_{2}$. So, this possibility cannot happen.
\end{itemize}
This finishes the proof for $\CC_{top}$. \qed
\end{proof}

\begin{remark}
To frame the last result, it is interesting to note that no category
$\CC_{*}$ considered in this paper has all binary pushouts,
essentially because the category of sets with injective mappings as
morphisms does not have all binary pushouts, either.  As a matter of
fact, the simplest counter-example does not involve arcs at all.  Let
$S$ be the empty tree and, for every $i=1,2$, let $T_{i}$ be the tree
consisting of a single node $\{a_{i}\}$ and no arc, and let
$m_{i}:V(S)\to V(T_{i})$ be the empty mapping.  It is clear that each
$m_{i}$ is a $\CC_{*}$-embedding, for every category $\CC_{*}$.  Now,
assume that $m_{1}:S\to T_{1}$ and $m_{2}:S\to T_{2}$ have a pushout
$(P, g_{1}:T_{1}\to P,g_{2}:T_{2}\to P)$ in $\CC_{*}$.

Consider the tree $P'$ consisting of two nodes $a_{1},a_{2}$ and no
arc and the mappings $g'_{1}:V(T_{1})\to V(P')$ and
$g'_{2}:V(T_{2})\to V(P')$ defined by $g_{1}'(a_{1})=a_{1}$ and
$g_{2}'(a_{2})=a_{2}$.  These mappings are $\CC_{*}$-embeddings, for
every category $\CC_{*}$.
Since $g_{1}'\circ m_{1}=g_{2}'\circ m_{2}$, by the universal property
of pushouts there exists a $\CC_{*}$-embedding $g':P\to P'$ such that
$g'\circ g_{1}=g_{1}'$ and $g'\circ g_{2}=g_{2}'$: in particular,
$g'(g_{1}(a_{1}))=a_{1}\neq a_{2}=g'(g_{2}(a_{2}))$, and therefore
$g_{1}(a_{1})\neq g_{2}(a_{2})$.

Consider now the tree $P''$ consisting of a single node $a$ and no arc
and the mappings $g''_{1}:V(T_{1})\to V(P'')$ and $g''_{2}:V(T_{2})\to
V(P'')$ defined by $g_{1}''(a_{1})=g_{2}''(a_{2})=a$.  Again, these
mappings are $\CC_{*}$-embeddings, for every category $\CC_{*}$, and
they satisfy that $g_{1}''\circ m_{1}=g_{2}''\circ m_{2}$.  Then, by the
universal property of pushouts, there exists a $\CC_{*}$-embedding
$g'':P\to P''$ such that $g''\circ g_{1}=g_{1}''$ and $g''\circ
g_{2}=g_{2}''$.  But then
$g''(g_{1}(a_{1}))=g_{1}''(a_{1})=a=g_{2}''(a_{2})=g''(g_{2}(a_{2}))$,
and hence  $g''$ is not injective. Therefore, it cannot be a
$\CC_{*}$-embedding, which yields a contradiction.

This shows that $m_{1}$ and $m_{2}$ don't have a pushout in any
category $\CC_{*}$. Of course, in this case $S$ is not a least common
$\CC_{*}$-subtree of $T_{1}$ and $T_{2}$.
\end{remark}

\section{Largest common subtrees and smallest common supertrees}
\label{sec-frame}

Let $\CC_{*}$ still denote any category $\CC_{iso}$, $\CC_{hom}$,
$\CC_{top}$ or $\CC_{min}$.  In this section, we show that the
constructions presented in the last two sections can be used to
obtain largest common $\CC_{*}$-subtrees and smallest common
$\CC_{*}$-su\-per\-trees of pairs of trees.  The key will be the  
following
result.

\begin{lemma} \label{fitagral}
Let $T_{1}$ and $T_{2}$ be two trees, and let $T_{\mu}$ be a largest
common $\CC_{*}$-subtree of them.  For every common
$\CC_{*}$-su\-per\-tree $T$ of $T_{1}$ and $T_{2}$, we have that
$|V(T)|\geq |V(T_{1})|+|V(T_{2})|-|V(T_{\mu})|$.
\end{lemma}

\begin{proof}
Propositions~\ref{cond-I-min}, \ref{pb-top}, \ref{pb-hom}, and
\ref{pb-iso} show that, for every two $\CC_{*}$-embeddings
$f_{1}:T_{1}\to T$ and $f_{2}:T_{2}\to T$, there exists a common
$\CC_{*}$-subtree $T_{0}$ of $T_{1}$ and $T_{2}$ with set of nodes
containing $f_{1}(V(T_{1}))\cap f_{2}(V(T_{2}))$: after a relabeling
of the nodes (so that $f_{1}$ and $f_{2}$ are given by inclusions of
the sets of nodes), it will be the intersection $T_{p}$ of $T_{1}$ and
$T_{2}$ in $\CC_{iso}$, $\CC_{hom}$, and $\CC_{top}$, and its one-node
extension $\tilde{T}_{p}$ in $\CC_{min}$.  Then,
$$
|f_{1}(V(T_{1}))\cap f_{2}(V(T_{2}))| \leq |V(T_{0})| \leq
|V(T_{\mu})|
$$
and hence,
$$
\begin{array}{rl}
|V(T)| & \geq |f_{1}(V(T_{1}))\cup f_{2}(V(T_{2}))|\\
& = |f_{1}(V(T_{1}))|+|f_{2}(V(T_{2}))|-|f_{1}(V(T_{1}))
\cap f_{2}(V(T_{2}))|\\
&\geq  |V(T_{1})|+|V(T_{2})|-|V(T_{0})| \\
& \geq |V(T_{1})|+|V(T_{2})|-|V(T_{\mu})|,
\end{array}
$$
as we claimed. \qed
\end{proof}

\begin{theorem} \label{sm-sup-po}
For every pair of trees $T_{1}$ and $T_{2}$, any $\CC_{*}$-sum of
$T_{1}$ and $T_{2}$ is a smallest common $\CC_{*}$-su\-per\-tree of  
them.
\end{theorem}

\begin{proof}
By Proposition~\ref{th-supertree-gral}, any $\CC_{*}$-sum $T_{\sigma}$
of $T_{1}$ and $T_{2}$ is a common $\CC_{*}$-su\-per\-tree of them, and  
by
construction $$|V(T_{\sigma})|= |V(T_{1})|+|V(T_{2})|-|V(T_{\mu})|,$$
for some largest common $\CC_{*}$-subtree $T_{\mu}$ of them.  Thus,
$T_{\sigma}$ achieves the lower bound established in Lemma
\ref{fitagral} for common $\CC_{*}$-su\-per\-trees of $T_{1}$ and  
$T_{2}$,
which implies that it is a smallest common $\CC_{*}$-su\-per\-tree of
them.  \qed
\end{proof}

\begin{theorem} \label{gr-sub-pb}
For every two trees $T_{1}$ and $T_{2}$, any intersection of $T_{1}$
and $T_{2}$ obtained through $\CC_{*}$-embeddings into a smallest
common $\CC_{*}$-su\-per\-tree of them is a largest common
$\CC_{*}$-subtree of $T_{1}$ and $T_{2}$.
\end{theorem}

\begin{proof}
Let $T_{1}$ and $T_{2}$ be two trees, let $T'_{\sigma}$ be a smallest
common $\CC_{*}$-su\-per\-tree of $T_{1}$ and $T_{2}$, let  
$p_{1}:T_{1}\to
T'_{\sigma}$ and $p_{2}:T_{2}\to T'_{\sigma}$ be any
$\CC_{*}$-embeddings, and let $T'_{p}$ be any common
$\CC_{*}$-subtree of $T_{1}$ and $T_{2}$ obtained by expanding the
intersection $T_{p}$ of $T_{1}$ and $T_{2}$ obtained through $p_{1}$
and $p_{2}$, which exists by Propositions~\ref{cond-I-min},
\ref{pb-top}, \ref{pb-hom}, and \ref{pb-iso}.

Now, by Theorem~\ref{sm-sup-po} we have that, for any largest common
$\CC_{*}$-subtree $T_{\mu}$ of $T_{1}$ and $T_{2}$,
$$|V(T'_{\sigma})|=|V(T_{1})|+|V(T_{2})|-|V(T_{\mu})|$$ and we know
that $$|p_{1}(V(T_{1}))\cap p_{2}(V(T_{2}))|\leq |V(T'_{p})|\leq
|V(T_{\mu})|.$$
Then,
$$
\begin{array}{l}
|V(T_{1})|+|V(T_{2})|-|V(T_{\mu})|  \\
\qquad\qquad = |V(T'_{\sigma})| \geq |p_{1}(V(T_{1}))\cup  
p_{2}(V(T_{2}))|\\
\qquad\qquad = |p_{1}(V(T_{1}))|+|p_{2}(V(T_{2}))|-|p_{1}(V(T_{1}))\cap  
p_{2}(V(T_{2}))|\\
\qquad\qquad \geq |V(T_{1})|+|V(T_{2})|-|V(T'_{p})|\\
\qquad\qquad \geq |V(T_{1})|+|V(T_{2})|-|V(T_{\mu})|.
\end{array}
$$
This implies that $|V(T'_{p})|= |V(T_{\mu})|= |p_{1}(V(T_{1}))\cap
p_{2}(V(T_{2}))|$.  From these equalities we deduce, on the one hand,
that $T'_{p}$ is also a largest common $\CC_{*}$-subtree of $T_{1}$
and $T_{2}$, and on the other hand, that $V(T'_{p})=
p_{1}(V(T_{1}))\cap p_{2}(V(T_{2}))$, i.e., that $T'_{p}=T_{p}$, as we
claimed.  \qed
\end{proof}

Thus, for every pair of trees $T_{1}$ and $T_{2}$, the pushout in
$\CC_{*}$ of any $\CC_{*}$-embeddings from a largest common
$\CC_{*}$-subtree of them yields a smallest common $\CC_{*}$-supertree
of them, and the pullback in $\CC_{*}$ of any $\CC_{*}$-embeddings
into a smallest common $\CC_{*}$-su\-per\-tree of them yields a
largest common $\CC_{*}$-subtree of them.  Moreover, all smallest
common $\CC_{*}$-super\-trees and all largest common $\CC_{*}$-subtrees
are obtained in this way up to isomorphisms, as the following corollaries show.

\begin{corollary} \label{cor-sm-sup-po}
Every smallest common $\CC_{*}$-supertree of a pair of trees $T_{1}$
and $T_{2}$ is, up to an isomorphism, the $\CC_{*}$-sum of $T_{1}$ and
$T_{2}$ obtained through the embeddings of a largest common
$\CC_{*}$-subtree into them.
\end{corollary}

\begin{proof}
Let $T_{1}$ and $T_{2}$ be two trees, let $T'_{\sigma}$ be a smallest
common $\CC_{*}$-supertree of $T_{1}$ and $T_{2}$ and let
$p_{1}:T_{1}\to T'_{\sigma}$ and $p_{2}:T_{2}\to T'_{\sigma}$ be any
$\CC_{*}$-embeddings.  By Theorem \ref{gr-sub-pb}, the intersection
$T_{p}$ of $T_{1}$ and $T_{2}$ obtained through $p_{1}$ and $p_{2}$,
together with the corresponding inclusions $\iota_{1}:T_{p}\to T_{1}$
and $\iota_{2}:T_{p}\to T_{2}$, is a largest common $\CC_{*}$-subtree
of $T_{1}$ and $T_{2}$.  Let now $T_{\sigma}$, together with
$m_{1}:T_{1}\to T_{\sigma}$ and $m_{2}:T_{2}\to T_{\sigma}$, be the
sum of $T_{1}$ and $T_{2}$ obtained through $\iota_{1}$ and
$\iota_{2}$.  By Theorem \ref{sm-sup-po}, $T_{\sigma}$ is a smallest
common $\CC_{*}$-supertree of $T_{1}$ and $T_{2}$, and by Theorem
\ref{th-po}, $(T_{\sigma},m_{1}:T_{1}\to T_{\sigma},m_{2}:T_{2}\to
T_{\sigma})$ is a pushout of $\iota_{1}:T_{p}\to T_{1}$ and
$\iota_{2}:T_{p}\to T_{2}$ in $\CC_{*}$.  Since $p_{1}\circ
\iota_{1}=p_{2}\circ \iota_{2}$, by the universal property of pushouts
there exists a $\CC_{*}$-embedding $p:T_{\sigma}\to T'_{\sigma}$ such
that $p\circ m_{1}=p_{1}$ and $p\circ m_{2}=p_{2}$.  Now, $T_{\sigma}$
and $T'_{\sigma}$ have the same size, because they are both smallest
common $\CC_{*}$-supertrees of $T_{1}$ and $T_{2}$. Therefore, 
$p:T_{\sigma}\to T'_{\sigma}$ is bijective, and thus an 
isomorphism by Lemma \ref{lem-iso}.
\qed
\end{proof}

A similar argument, which we leave to the reader, proves also the 
following result.

\begin{corollary} \label{cor-gr-sub-pb}
Every largest common $\CC_{*}$-supertree of a pair of trees $T_{1}$
and $T_{2}$ is, up to an isomorphism, the intersection of $T_{1}$ and
$T_{2}$ obtained through their embeddings into a smallest common
$\CC_{*}$-supertree. 
\end{corollary}

\begin{corollary} \label{cor-lin-equiv}
The problems of finding a largest common $\CC_{*}$-subtree and a
smallest common $\CC_{*}$-su\-per\-tree of two trees, in each case
together with a pair of witness $\CC_{*}$-embed\-dings, are reducible to
each other in time linear in the size of the trees.
\end{corollary}

\begin{proof}
Given two trees $T_{1}$ and $T_{2}$, if we know a largest common
$\CC_{*}$-subtree $T_{\mu}$ of them, together with a
pair of witness $\CC_{*}$-embeddings $m_{1}:T_{\mu}\to
T_{1}$ and $m_{2}:T_{\mu}\to T_{2}$, then the
construction of the pushout
$$
(T_{\sigma}, \ell_{1}:T_{1}\to T_{\sigma},
\ell_{2}:T_{2}\to T_{\sigma})
$$
of $m_{1}$ and $m_{2}$ described in Theorem~\ref{th-po} gives a
smallest common $\CC_{*}$-su\-per\-tree of $T_{1}$ and $T_{2}$, and
this construction can be obtained in time linear in the size of $T_1$
and $T_2$, as follows.

First, make copies $T'_1$ and $T'_2$ of $T_1$ and $T_2$, with
$\ell_1:T_1 \to T'_1$ and $\ell_2:T_2 \to T'_2$ identity mappings.
Second, sum up $T'_1$ and $T'_2$ into a graph $T_{\sigma}$. Third, for
each $a \in V(T_\mu)$, merge nodes $\ell_1(m_1(a))$ and
$\ell_2(m_2(a))$, and remove all parallel arcs.

Next, remove from $T_{\sigma}$ all arcs subsumed by paths, as follows.
For each node $y \in V(T_{\sigma})$ of in-degree 2, let $x,x' \in
V(T_{\sigma})$ be the source nodes of the two arcs coming into $y$.
Now, perform a simultaneous traversal of the paths of arcs coming into
$x$ and $x'$, until reaching node $x'$ along the first path or $x$
along the second path. The simultaneous traversal of incoming paths
may stop along either path, but continue along the other one, because
a node of in-degree 0 or in-degree 2 is reached. Finally, remove from
$T_{\sigma}$ either arc $(x',y)$, if node $x'$ was reached along the
first path, or arc $(x,y)$, if node $x$ was reached along the second
path.

Conversely, if we know a smallest common $\CC_{*}$-su\-per\-tree $T$ of
$T_{1}$ and $T_{2}$, together with a pair of witness
$\CC_{*}$-embeddings $f_{1}:T_{1}\to T$ and $f_{2}:T_{2}\to T$, then,
by Theorem~\ref{gr-sub-pb}, the pullback $$(T_{p}, \iota_{1}:T_{p}\to
T_{1},\iota_{2}:T_{p}\to T_{2})$$ of $f_{1}$ and $f_{2}$ described in
Section~\ref{sec-pb} yields a largest common $\CC_{*}$-subtree of
$T_{1}$ and $T_{2}$, and this construction can also be obtained in
time linear in the size of $T_1$ and $T_2$, as follows.

First, make a copy $T_p$ of $T$, with $g:T \to T_p$ the identity
mapping.  Second, for each $a \in V(T_1)$, mark $g(f_1(a))$ in $T_p$.
Third, for each $a \in V(T_2)$, if $g(f_2(a))$ is already marked in
$T_p$, double-mark it.  Next, for each node of $T_p$ which is not
double-marked, add a new arc from its parent (if any) to each of its
children (if any) in $T_p$, and remove the node not double-marked.
Finally, set mappings $\iota_i:T_p \to T_i$ for $i=1,2$, as follows:
for each $a \in V(T_i)$, if $g(f_i(a))$ is defined, set
$\iota_i(g(f_i(a)))=a$.
\qed
\end{proof}

\section{Conclusion} \label{sec-concl}

Subtree isomorphism and the related problems of largest common subtree
and smallest common supertree belong to the most widely used
techniques for comparing tree-structured data, with practical
applications in combinatorial pattern matching, pattern recognition,
chemical structure search, computational molecular biology, and other
areas of engineering and life sciences. Four different embedding
relations are of interest in these application areas: isomorphic,
homeomorphic, topological, and minor embeddings.

The complexity of the largest common subtree problem and the smallest
common supertree problem under these embedding relations is already
settled: they are polynomial-time solvable for isomorphic,
homeomorphic, and topological embeddings, and they are NP-complete for
minor embeddings. Moreover, efficient algorithms are known for largest
common subtree under isomorphic, homeomorphic, and topological
embeddings, and for smallest common supertree under isomorphic and
topological embeddings, and an exponential algorithm is known for
largest common subtree under minor embeddings.

In this paper, we have established the relationship between the
largest common subtree and the smallest common supertree of two trees
by means of simple constructions, which allow one to obtain the
largest common subtree from the smallest common supertree, and vice
versa. We have given these constructions for isomorphic, homeomorphic,
topological, and minor embeddings, and have shown their implementation
in time linear in the size of the trees.
In doing so, we have filled the gap by providing a simple extension of
previous largest common subtree algorithms for solving the smallest
common supertree problem, in particular under homeomorphic and minor
embeddings for which no  algorithm has been known previously.

Beside the practical interest of these extensions to previous
algorithms, we have provided a unified algebraic construction showing the
relation between largest common subtrees and smallest common
supertrees for the four different embedding problems studied in the
literature: isomorphic, homeomorphic, topological, and minor
embeddings. The unified construction shows that smallest common
supertrees are pushouts and largest common subtrees are pullbacks.

\textbf{Acknowledgements.} F. Rossell\'o was partially supported by
the Spanish DGES and the EU program FEDER, project ALBIOM
(BFM2003-00771).  G. Valiente was partially supported by Spanish CICYT
projects MAVERISH (TIC2001-2476-C03-01) and GRAMMARS
(TIN2004-07925-C03-01), and by the Ministry of Education, Science,
Sports and Culture of Japan through Grant-in-Aid for Scientific
Research B-15300003 for visiting JAIST (Japan Advanced Institute of
Science and Technology).  The authors acknowledge with thanks the
anonymous referees, whose comments, suggestions and criticism have led
to a substantial improvement of this paper.


\begin{thebibliography}{10}

\bibitem{amir.keselman:1997}
A.~Amir, D.~Keselman, Maximum agreement subtree in a set of
evolutionary trees: Metrics and efficient algorithms, SIAM Journal on
Computing 26~(6) (1997) 1656--1669.


\bibitem{aoki.ea:2003}
K.~F. Aoki, A.~Yamaguchi, Y.~Okuno, T.~Akutsu, N.~Ueda, M.~Kanehisa,
H.~Mamitsuka, Efficient tree-matching methods for accurate
carbohydrate database queries, in: Proc.\ 14th Int.\ Conf.\ Genome
Informatics, Universal Academy Press, 2003, pp.  134--143.

\bibitem{artymiuk.ea:2005}
P.~J. Artymiuk, R.~V. Spriggs, P.~Willett, Graph theoretic methods for
the analysis of structural relationships in biological macromolecules,
Journal of the American Society for Information Science and Technology
56~(5) (2005) 518--528.

\bibitem{barnard:1993}
J.~M. Barnard, Substructure searching methods: Old and new, Journal of
Chemical Information and Computer Sciences 33 (1993) 532--538.

\bibitem{chung:1987}
M.-J. Chung, ${O}(n^{2.5})$ time algorithms for the subgraph
homeomorphism problem on trees, Journal of Algorithms 8~(1) (1987)
106--112.

\bibitem{cole.ea:2000}
R.~Cole, M.~Farach-Colton, R.~Hariharan, T.~M. Przytycka, M.~Thorup,
An ${O}(n \log n)$ algorithm for the maximum agreement subtree problem
for binary trees, SIAM Journal on Computing 30~(5) (2000) 1385--1404.

\bibitem{conte.ea:2004}
D.~Conte, P.~Foggia, C.~Sansone, M.~Vento, Thirty years of graph
matching in pattern recognition, Int.\ J.\ Pattern Recogn.\ Artificial
Intell.  18~(3) (2004) 265--298.

\bibitem{dessmark.ea:2000}
A.~Dessmark, A.~Lingas, A.~Proskurowski, Faster algorithms for
subgraph isomorphism of $k$-connected partial $k$-trees, Algorithmica
27~(1) (2000) 337--347.

\bibitem{dulucq.tichit:2003}
S.~Dulucq, L.~Tichit, {RNA} secondary structure comparison: Exact
analysis of the {Z}hang-{S}hasha tree edit algorithm, Theoretical
Computer Science 306~(1--3) (2003) 471--484.

\bibitem{fernandez.valiente:2001}
M.-L. Fern\'andez, G.~Valiente, A graph distance measure combining
maximum common subgraph and minimum common supergraph, Pattern
Recognition Letters 22~(6--7) (2001) 753--758.

\bibitem{gillet.ea:2003}
V.~J. Gillet, P.~Willett, J.~Bradshaw, Similarity searching using
reduced graphs, Journal of Chemical Information and Computer Sciences
43~(2) (2003) 338--345.


\bibitem{gupta.nishimura:1998}
A.~Gupta, N.~Nishimura, Finding largest subtrees and smallest
supertrees, Algorithmica 21~(2) (1998) 183--210.

\bibitem{jansson.lingas:2003}
J.~Jansson, A.~Lingas, A fast algorithm for optimal alignment between
similar ordered trees, Fundamenta Informaticae 56~(1--2) (2003)
105--120.

\bibitem{jiang.ea:jda:2004}
T.~Jiang, G.~Lin, B.~Ma, K.~Zhang, The longest common subsequence
problem for arc-annotated sequences, Journal of Discrete Algorithms
2~(2) (2004) 257--270.

\bibitem{jiang.ea:1995}
T.~Jiang, L.~Wang, K.~Zhang, Alignment of trees---an alternative to
tree edit, Theoretical Computer Science 143~(1) (1995) 137--148.


\bibitem{kilpeleinen.manila:1995}
P.~Kilpel{\"a}inen, H.~Mannila, Ordered and unordered tree inclusion,
SIAM Journal on Computing 24~(2) (1995) 340--356.

\bibitem{matousek.thomas:1992}
J.~Matou{\v{s}}ek, R.~Thomas, On the complexity of finding
isomorphisms and other morphisms for partial $k$-trees, Discrete
Mathematics 108~(1--3) (1992) 343--364.

\bibitem{nishimura.ea:2000}
N.~Nishimura, P.~Ragde, D.~M. Thilikos, Finding smallest supertrees
under minor containment, Int.\ Journal of Foundations of Computer
Science 11~(3) (2000) 445--465.

\bibitem{pinter.ea:2004}
R.~Y. Pinter, O.~Rokhlenko, D.~Tsur, M.~Ziv-Ukelson, Approximate
labelled subtree homeomorphism, in: Proc.\ 15th Annual Symp.\
Combinatorial Pattern Matching, Vol.  3109 of Lecture Notes in
Computer Science, Springer-Verlag, 2004, pp.  55--69.


\bibitem{pinter.ea:2005}
R.~Y. Pinter, O.~Rokhlenko, E.~Yeger-Lotem, M.~Ziv-Ukelson, Alignment
of metabolic pathways, Bioinformatics 21~(16) (2005) 3401--3408.


\bibitem{shamir.tsur:1999}
R.~Shamir, D.~Tsur, Faster subtree isomorphism, Journal of Algorithms
33~(2) (1999) 267--280.

\bibitem{shasha.wang.ea:1994}
D.~Shasha, J.~T.-L. Wang, K.~Zhang, F.~Y. Shih, Exact and approximate
algorithms for unordered tree matching, {IEEE} Transactions on
Systems, Man, and Cybernetics 24~(4) (1994) 668--678.

\bibitem{shasha.zhang:1990}
D.~Shasha, K.~Zhang, Fast algorithms for the unit cost editing
distance between trees, Journal of Algorithms 11~(4) (1990) 581--621.

\bibitem{steel.warnow:1993}
M.~A. Steel, T.~Warnow, Kaikoura tree theorems: Computing the maximum
agreement subtree, Information Processing Letters 48~(2) (1993)
77--82.

\bibitem{torsello.ea:2005}
A.~Torsello, D.~H. Rowe, M.~Pelillo, Polynomial-time metrics for
attributed trees, IEEE Trans.  Pattern Anal.  Mach.  Intell.  27~(7)
(2005) 1087--1099.

\bibitem{valiente:2002}
G.~Valiente, Algorithms on Trees and Graphs, Springer-Verlag, Berlin,
2002.

\bibitem{valiente:cpm.2003}
G.~Valiente, Constrained tree inclusion, in: Proc.\ 14th Annual Symp.\
Combinatorial Pattern Matching, Vol.  2676 of Lecture Notes in
Computer Science, Springer-Verlag, 2003, pp.  361--371.

\bibitem{valiente:cpm:jda}
G.~Valiente, Constrained tree inclusion, Journal of Discrete
Algorithms 3~(2--4) (2005) 431--447.

\bibitem{wang.zhao:2003}
L.~Wang, J.~Zhao, Parametric alignment of ordered trees,
Bioinformatics 19~(17) (2003) 2237--2245.

\bibitem{zhang:2005}
K.~Zhang, RNA structure comparison and alignment, in: J.~T.-L. Wang,
M.~J. Zaki, H.~Toivonen, D.~Shasha (Eds.), Data Mining in
Bioinformatics, Springer, 2005, pp.  59--81.

\bibitem{zhang.shasha:1989}
K.~Zhang, D.~Shasha, Simple fast algorithms for the editing distance
between trees and related problems, SIAM Journal on Computing 18~(6)
(1989) 1245--1262.





\end{thebibliography}
\end{document}